\documentclass[reprint,12pt,onecolumn,notitlepage,nofootinbib]{revtex4-1}

\usepackage{epsfig}
\usepackage{amsmath}   
\usepackage{amsfonts}
\usepackage{amssymb}
\usepackage{graphicx}  
\usepackage{verbatim}
\usepackage{color}   
\usepackage{subfigure}
\usepackage{hyperref}
\usepackage{cancel}
\newcommand{\citen}[1]{\cite{#1}}
\newcommand{\tbl}[1]{\caption{#1}}
\makeindex

\newcommand{\curlyP}{{\mathcal P}}
\newcommand{\curlyV}{{\mathcal V}}

\definecolor{matthieu}{RGB}{0,0,0}
\definecolor{jim}{RGB}{0,0,0}
\definecolor{albe}{RGB}{0,0,0}
\newcommand{\mw}[1]{\textcolor{matthieu}{#1}}
\newcommand{\jps}[1]{\textcolor{jim}{#1}}
\newcommand{\ar}[1]{\textcolor{albe}{#1}}

\newcommand{\Subsection}{\section}
\newcommand{\Subsubsection}{\subsection}

\usepackage{perpage}\MakePerPage{footnote}

\begin{document}

\title{Avalanches and deformation in glasses and disordered systems
\label{chap:Avalanches}\\
{\rm Contribution to the book {\em Spin Glass Theory and Far Beyond - Replica Symmetry Breaking after 40 Years}}}

\author{Alberto Rosso}

\affiliation{Universit\'{e} Paris Saclay, CNRS,LPTMS, 91405, Orsay, France}

\author{James P. Sethna}

\affiliation{LASSP, Cornell University, Ithaca, NY 14853, USA}

\author{Matthieu Wyart}

\affiliation{Institute of Physics, \'Ecole Polytechnique F\'ed\'erale de Lausanne (EPFL), CH-1015 Lausanne, Switzerland}


\begin{abstract}
In this chapter, we discuss avalanches in glasses and disordered 
systems, and the macroscopic dynamical behavior that they mediate.  We
briefly review three classes of systems where avalanches are observed:
depinning transition of disordered interfaces, yielding of amorphous materials,
and the jamming transition. Without extensive
formalism, we discuss results gleaned from theoretical approaches -- 
mean-field theory, scaling and exponent relations, the renormalization
group, and a few results from replica theory. We focus both on the
remarkably sophisticated physics of avalanches and on relatively new
approaches to the macroscopic flow behavior exhibited past the
depinning/yielding transition.
\end{abstract}

\maketitle
\Subsection{Introduction}\label{sec:AvalancheIntro}
 
Rigid systems, put under stress, will often respond through a series of avalanches~\cite{SethnaDM01,SethnaDP06,Zapperi22}
with a broad distribution of sizes and durations.%
  \footnote{Warning: Everyday avalanches in sandpiles, or of snow or rocks on mountains, rarely exhibit the fractal, scale invariant behavior we study here~\cite{NagelAngleOfRepose}.} Faults in
the Earth exhibit earthquakes spanning many decades of size~\cite{Fisher1997,DahmenEarthquakes06,jagla2014,dearcangelis2016} when stressed by the motion of tectonic plates. Iron placed in a growing magnetic field will emit Barkhausen crackling noise (both acoustic and electromagnetic)
as the internal magnetic domain walls shift between metastable
states~\cite{Sethna07Crossover,DurinZapperi06}. Raindrops on your windshield advance in bursts -- their leading edges, pinned by dirt, releasing as they grow and merge~\cite{Raindrops}. Avalanches also arise microscopically in systems that usually are viewed as responding smoothly to an external stress. Your fork, bending irreversibly when you try to cut through a tough piece of meat, responds through many dislocation avalanches with a distribution of sizes -- measurable, for example, in experiments where nanopillars of metal are compressed~\cite{Sethna07Wire,DahmenNanopillars,SethnaBDGGHKLLNQRRSZ17,NiZLMDSG19,PapanikolaouDCSUWZ12,Csikor07}. Toothpaste squeezed out of a tube yields above a critical stress through a jerky set of avalanches.

In all these systems an avalanche starts from a weak spot that becomes unstable and slips. This triggers other instabilities nearby that in turn infect other regions -- sometimes halting quickly, and sometimes sweeping over vast regions before halting. Some systems exhibit avalanches of all scales (sometimes termed `crackling noise') only when delicately tuned to balance between stability (only local events) and instability (continuous flow). We shall see that many systems can either {\em self-organize} to this balancing point, or exhibit {\em generic scale invariance} -- with events of all scales due to long-range non-monotonic ({i.e.} of varying sign) interactions.

These avalanches are characteristic of systems that have a complex energy landscape (well studied in other chapters of this book). Increasing the stress on the system `tilts' this landscape, and the avalanches represent transitions from one metastable valley to another. For a sufficiently large tilt, rigidity can be lost and the system flows continuously. We focus on unifying concepts that describe both the avalanche-type response and continuous flows when it occurs, in systems ranging from disordered magnets and granular materials to epidemics, raindrops, and brain activity. We shall explain, without extensive formalism, how {\em scaling} and {\em renormalization-group} methods unify all of these systems, and elucidate the important physics that make these systems different from one another.

Our contribution is organized as follows. In section~\ref{sec:systems}, we introduce a few of the many systems exhibiting collective avalanches and crackling noise: epidemics, depinning of lines and interfaces in random media, yielding in amorphous solids such as glasses and foams, and dense granular flows. In the first two cases, system-spanning avalanches occur at a critical point, while in other cases they occur in an entire phase. 
 In section~\ref{section:stability}, we explain these facts by introducing the notion of {\it excitations}, as regions of the material that are about to undergo an instability. If sufficiently long-range interactions are present, requiring stability of the material constrains the density of excitations, which in turn implies a generic scale invariance in the entire rigid phase.
 In sections~\ref{subsec:DepinningTheory}, \ref{subsec:YieldingTheory} and \ref{subsec:jammingtheory} we introduce renormalization group results and scaling arguments. They apply but differ in the systems considered, allowing one to both compare these systems and build a detailed understanding of several phenomena, including the shape of avalanches or their connection with stationary flows when rigidity is lost. In section~\ref{subsec:conclusions}, we conclude by emphasizing key open questions of this field.

\Subsection{Systems considered}\label{sec:systems}

Avalanches appear in many areas, both in science and engineering and
in other fields. In addition to the examples above,
avalanche models have been used to describe fracture
precursors in quasi-brittle material like bones and
seashells~\cite{ShekhawatZS13}, bubble avalanches in foams~\cite{TewariSDKLL99},
fluids invading porous media~\cite{CieplakR88, ortin2017}
(like milk invading puffed rice cereal~\cite{url:CracklingKids} or coffee soaking into a
napkin~\cite{AlavaCoffeeAvalanches}),
vortex avalanches
in superconductors~[\citen{Zapperi22}, Ch.~9] and neutron stars~\cite{StarQuakes},
wars~\cite{WarAvalanches}, neural avalanches in brain
tissue~\cite{DahmenNeurons,beggsP03}, noise while tearing paper~\cite{AlavaPaperTearing} and while crumpling paper or candy wrappers~\cite{HouleS96,KramerL96},
 and clapping after concert
performances~\cite{MichardB05}. In this section, we introduce four phenomena we focus on: the mean field theory of epidemics, the depinning transition of an elastic manifold in a disordered environment, the plasticity of simple yield stress materials, and the jamming transition in granular systems.

 \Subsubsection{Mean-field pandemic model}
\label{subsec:Pandemic}

Outbreaks of disease are a public-health manifestation of avalanche behavior.
An unlucky person is infected by a bird, pig, or bat, and infects one or more
surrounding people. Let $R_0$ denote the number of infections triggered
by each formerly infected person; $R_0 \sim 12{-}18$ for measles, $2{-}3$ for
influenza. For a new disease where nobody is immune, $R_0<1$ means that the
disease will gradually disappear, $R_0>1$ implies that -- if not stamped out
early -- it will cause a global pandemic. Near $R_0=1$, there is a phase
transition, with avalanches (outbreaks that halt) on all scales for $R_0 \lesssim 1$.

This model is called the
Bienaym\'e-Galton-Watson process~\cite{bienayme1845,watson1875}. It is a
{\em mean-field} model for avalanches, because there is no notion of space: everyone can infect anyone. It also describes a fully connected Ising model in a
disordered material~\cite{SethnaDKKRS93}, where a spin flips when its external
field increases beyond its random threshold, increasing the external
`mean field' on all the other $N$ spins enough to on average flip $R_0$
neighbors. In some cases, this mean-field aspect is not a realistic assumption: plants do not move, so crop diseases that spread locally have avalanches that are correlated in space.

Consider the number $I_{N_R}$ of infected people at the time when $N_R$
people have stopped being infectious (either recovered or deceased). Each
person, before they recover, infects $\xi_{N_R}$ others, where
$\xi$ is a random integer chosen from a Poisson distribution of mean $R_0 I_{N_R}$.
For simplicity, we assume these infections happen at the time when the
person recovers, implying $I_{N_R+1}=I_{N_R}+\xi_{N_R}$. This random walk
halts at pandemic size $S$ when $I_S$ first touches zero; its size $S$ is the number of random steps for the first return to the origin.
The probability of a pandemic of size $S$ at $R_0=1$ thus coincides with the probability that a random walk starting at the origin will have its first return after $S$ steps, which can be calculated to be a power law
$P(S) \sim
S^{-\tau}$ with $\tau = 3/2$. Let us measure the distance to the
critical point as $r = (R_0-R)/R$. For $r \lesssim 0$ below threshold,
the random walk starts positive ($I_1=1$) but has a negative drift -- it is a biased random walk. For small avalanches near $R_c$, the accumulated drift $\Delta I = r S$ is negligible compared to the typical random walk distance $I_\text{max} \sim S^{1/2}$, but for large avalanches the drift dominates, making large avalanches unusual. Equating the two, we expect the probability distribution of avalanches $P(S,r)$ is cut off at a size $S_\text{max} \sim r^{-1/\sigma}$ with $\sigma = 1/2$. More precisely, $P(S,r)$ can be shown to follow
\begin{equation}
\label{eq:curlyPMeanField}
P(S,r) 	\sim S^{-3/2} \exp(-S r^2/2)
	\sim S^{-\tau} \curlyP(S^\sigma r) ~~~~~ \mathrm{with}\ \
\curlyP(X) = \exp(-X^2/2) \ .
\end{equation}
Here $\tau$ and $\sigma$ are universal critical exponents, and $\curlyP$
is a universal scaling function of the invariant scaling ratio
$X=S^\sigma r$. Universality signifies that changing our epidemic model
to a more realistic stochastic SIR model (a `compartmental' model discussing the susceptible,
infected, and recovered populations), or a model with superspreaders,
does not change the values for $\tau$, $\sigma$, and $\curlyP$ that describe large epidemics. However, some features like cities can change the behavior away from
mean-field in qualitative ways.

Scaling functions arise~\cite{Sethna22} whenever more than two parameters and properties are involved (here, probability $P$, depending on size $S$, and distance $r$ to the critical point). Other examples include avalanche duration versus time as a function of $r$, scaling depending on the system size $L$, crossover scaling~\cite{Sethna07Crossover,ChenZS15}, singular corrections to scaling, etc.
We shall study a universal scaling function for the average temporal shape of avalanches, which has two properties depending on two parameters, in section~\ref{subsubsec:AvalancheTemporalShape}.

Finally, note that to obtain avalanches of all scales requires to
tune a parameter like $R_0$ to a critical point. However, in various cases this parameter can spontaneously evolve to that value, a phenomenon called
{\em self organized criticality}~\cite{bak1987self}. In our disease outbreak model, imagine
we add a term which describes the societal response to widespread illness
-- wearing masks, getting immunized, avoiding large gatherings. As outbreaks
become large and alarming, these responses will tend to reduce the effective
interactions, tuning $R_0$ down until just below the threshold for a pandemic~\cite{ZapperiPandemicSOC}.

 \Subsubsection{Depinning transitions}
 \label{subsec:Depinning}

\begin{figure}
\centerline{\includegraphics[width=12cm]{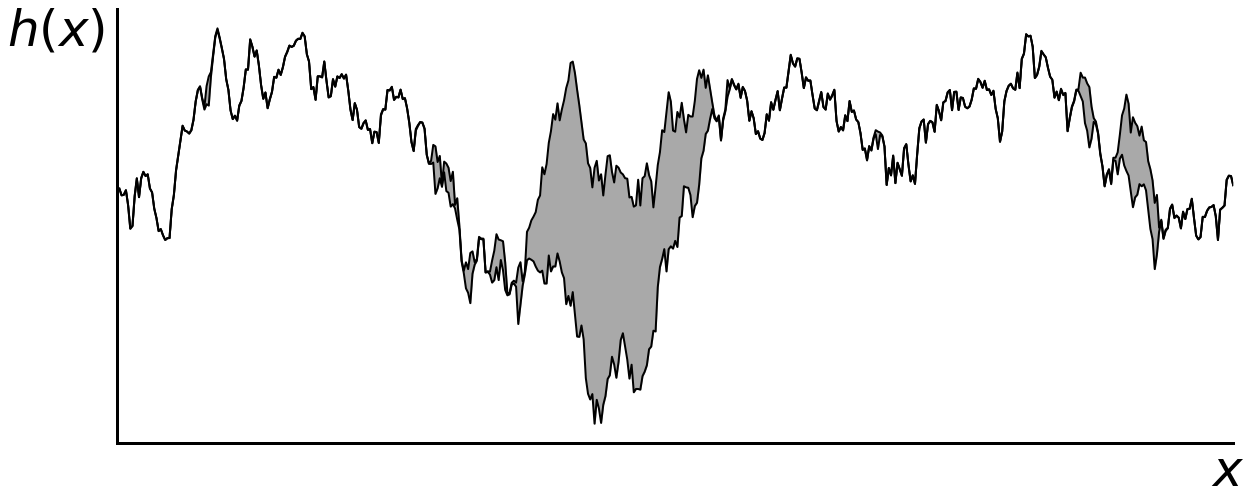}}
\caption{{\bf Depinning} of a one-dimensional manifold in a two dimensional environment. An increasing external field causes the manifold to locally destabilize, moving from one configuration to another in an avalanche (large gray region, of area $S$). If there are long-range forces in the problem, this event may cause other distant segments to destabilize as well (small gray regions to right).
}
\label{fig:Depinning}
\end{figure}

View Fig.~\ref{fig:Depinning} as a \ar{one}-dimensional elastic interface, pushed through a disorder environment by applying a force per unit length $f$. Examples of this situation abound%
\footnote{One can also view it as a two-dimensional disordered crystal
 with a jagged dislocation line pinned on impurities, or a polymer pinned
 on a disordered surface, or an oil-water interface in porous rock~\cite{CieplakR88}.}: it could represent a magnetic film, magnetized `up' below the jagged interface, and `down' above the interface with a forcing that depends on the applied magnetic field~\cite{DurinZapperi06}; or the triple line formed at the edge of a drop on your windshield, forced by gravity~\cite{Raindrops}.
These phenomena are well described as the propagation of a single-valued front \ar{$u(r,t)$, where $r$} and $u$ are the horizontal and vertical
coordinates of the interface. Neglecting inertial effects, the equation of motion of our front might be modeled as a disordered partial differential equation of this class~\cite{Fisher98, kardar1998}:
 \begin{equation}
 \label{eq:depinning}
     \partial_t u(r,t) = f_{\text{el}}\left[u\right]+ f_{\text{dis}}\left[u(r,t), x\right] +f \ .
 \end{equation}
Here $f_\text{el}$ is the elastic force trying to restore the straight interface, $f_\text{dis}$ is the force due to the random potential, and $f$ is the external drive.
The solution of this class of equations shows a continuous phase
transition with the velocity playing the role of the order parameter and
the force acting as the control parameter~\cite{larkin1970,fisher1985,middleton1992a,giamarchi2006,joerg2021theory}. In particular, below a
critical {\it depinning} force $f_\text{c}$ the steady velocity is zero.

{\it Avalanches:} If we
increase that external field for $f<f_\text{c}$, we can destabilize a weakly-pinned region,
triggering an avalanche (see Fig.~\ref{fig:Depinning}). Just as in the
epidemic model, below $f_\text{c}$ these avalanches have a size \jps{distribution} 
$P(S,f) \sim S^{-\tau} \mathcal{P}(S (f_\text{c}-f)^{1/\sigma})$, becoming scale free
when the force approaches $f=f_\text{c}$.
Individual avalanches smaller than the cutoff
$S_\text{max} = (f_\text{c}-f)^{1/\sigma}$ will spread over a spatial extent
$\sim S^{1/d_f}$, where $d_f=1/(\sigma \nu)$. They will have
a duration in time $\sim S^{z/d_f}$ that goes as their spatial extent
to the power $z$. Avalanche distributions have been extensively studied in the literature~\cite{alessandro1990,colaiori2008,ledoussal2009c,ledoussal2009b,rosso2009,le2013avalanche,delorme2016}.

{\it Flow properties are singular:} For $f>f_\text{c}$,
the interface will begin moving with a mean
velocity $v\sim (f-f_\text{c})^\beta$ in a jerky fashion,
with correlated jumps that mimic the avalanches found just below $f_\text{c}$.
Similarly to equilibrium continuous phase transitions, the collective
nature of this intermittent dynamics arises from the existence of a
correlation length $\xi\sim |f-f_\text{c}|^{-\nu}$ that diverges when the force
approaches $f_\text{c}$.

{\it Fractal structure:} At the depinning threshold, the interface is fractal, with a roughness
that is characterized by a typical vertical height change
$u(r,t)-u(r',t) \sim (r-r')^\zeta$ corresponding to a height-height
correlation function $C(r) = \langle (u(r')-u(r'+r))^2\rangle \sim r^{2 \zeta}$.

The elastic force $f_\text{el}$ on a domain wall can involve long-range
interactions~\cite{Zoia07}. This is the case for the triple line of a raindrop on
glass~\cite{joanny1984,Raindrops} (and also of crack fronts propagating in
brittle materials~\cite{gao1989,alava2006}).
The elastic force on the droplet
interface must include the surface tension energy of the wiggly drop
shape imposed by the jagged air-water interface \ar{$u(r)$, hence
$f_\text{el} \sim \int d^D r' (u(r')-u(r))/(r-r')^{D+\alpha}$,
with $\alpha =1$ and  $D=1$~\cite{joanny1984,gao1989,bonamy2008,ledoussal2006,bonamy2011}.
A second important example is provided by the sliding of frictional interfaces  or of localized shear bands in amorphous materials \cite{dahmen2009micromechanical,antonaglia2014bulk,wright2016experimental,wright2018slip}.  Here, if inertia, velocity weakening or visco-elastic effects can be neglected, the dynamics is depinning-like with $D=2$ and $\alpha=1$.  }

 The presence of long range elasticity does not change the qualitative picture of depinning given above, but for $\alpha < 2$ the value of the exponents $\beta, \zeta, \tau$ and $\sigma$ are modified; for $\alpha \ge 2$ short range exponents remain correct. Also, long range interactions can trigger distant weak spots to yield, leading to avalanches with disconnected pieces (see Fig.~\ref{fig:Depinning}), called clusters. Their sizes, distances and extensions display power law behaviors as for the full avalanche, but with new exponents.

Note that many physical systems self-organize at their
depinning transitions -- naturally exhibiting an emergent scale invariant
avalanche behavior and fractal front geometries. For example, in magnets, the weak
but long-range magnetic dipole fields lead to {\em demagnetizing}
forces~\cite[Sec.~8.2]{Zapperi22} that decrease the effective driving
force as the front advances.
Theoretical depinning models add an extra parabolic term to
Eq.~\eqref{eq:depinning} in order to model this self-organization.

In section~\ref{subsec:DepinningTheory} we will explain the relationship between avalanches below $f_\text{c}$, stationary flow above $f_\text{c}$, and the fractal structure of the interface at $f_\text{c}$, and how these properties are affected by long-range interactions.

 \Subsubsection{Plasticity in amorphous materials}
 \label{subsec:Plasticity}

Other disordered systems display out of equilibrium phase transitions induced by an external drive. For example, a yielding transition is observed in foams, emulsions and metallic glasses when a stress, $\Sigma$, is applied. For a small stress these materials deform as solids, but for stresses $\Sigma$ above a yield stress $\Sigma_\text{c}$ they flow as liquids. In the liquid phase, the strain rate vanishes non-linearly~\cite{Herschel1926,ovarlez2013rheopexy}, $\dot \epsilon \sim (\Sigma -\Sigma_\text{c} )^\beta$. This behavior is similar to depinning (albeit with a larger flow exponent $\beta$), but here the system exhibits generic scale invariance. Scale invariant avalanches start well below $\Sigma_\text{c}$.

\begin{figure}[t]
\centering
\includegraphics[
width= \linewidth]
{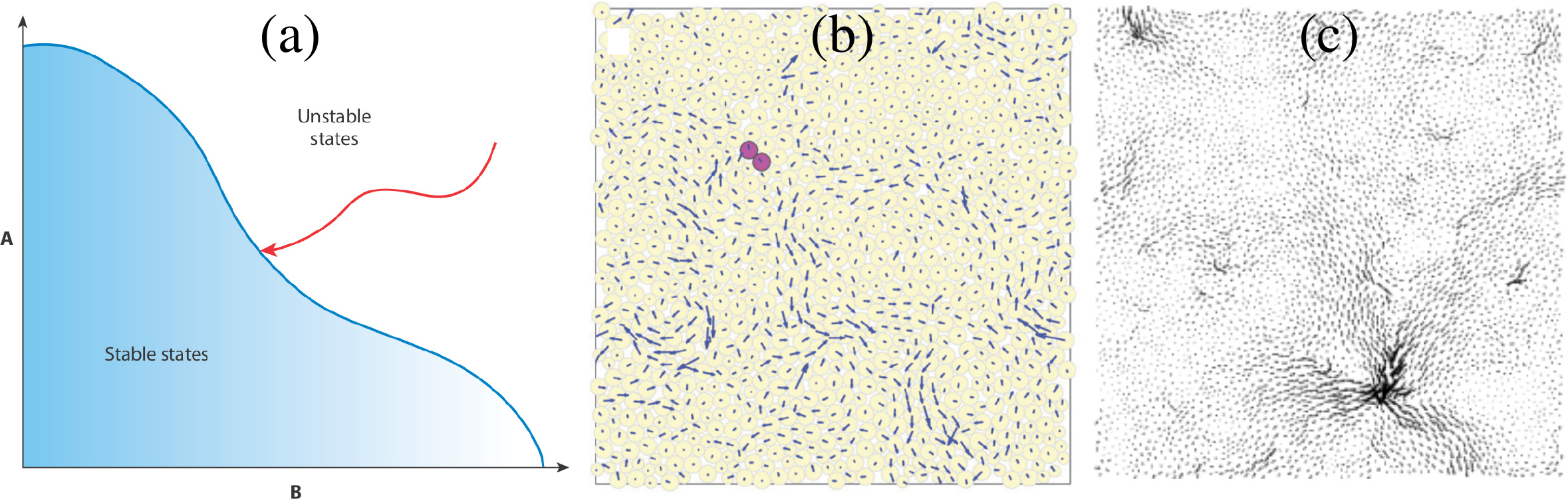}
 \caption{(a) Schematic stability diagram in configuration space. $A$ and $B$ are observables characterizing the configurations visited. The full black line corresponds to marginal stability: It separates regions in which excitations are stable and unstable, respectively. The arrow illustrates a dynamical trajectory of a system that is cooling from an initial high temperature phase by stepwise relaxation of individual elementary excitations. When the system reaches the marginality line, the excitations become stable. If these excitations are the main drive of the dynamics, the system slows down very rapidly as it enters into the stable region and freezes very close to the marginal stability line.  
 There, low-energy excitations are abundant, and rich dynamics, such as crackling noise, can occur. From Ref.~\cite{Muller14}. (b) Elementary excitation corresponding to the opening of a single contact in a packing of hard discs. From Ref.~\cite{Lerner13a}. (c) A shear transformation in an amorphous solid made of soft compressed particles. From Ref.~\cite{Maloney06a}.}
\label{fig:excitations}
\end{figure}

In the solid phase one observes plastic instabilities, called shear transformations~\cite{Argon79}, involving irreversible rearrangements of few bubbles or droplets. As illustrated in Fig.~\ref{fig:excitations}C, this local rearrangement induces a large, complex nearby rearrangement and a far-field power-law decay. This stress redistribution then can trigger other shear transformation zones that are kicked above their stability threshold.

Thus, as in depinning, this first instability can be the epicenter of an
avalanche~\cite{Maloney04,Salerno12}. The connection with the depinning
transition can be made explicit considering the growth of local strain
$\epsilon(r,t)$ at a given stress $\Sigma$. Its evolution can be
written~\cite{Lin14} in the form of Eq.~(\ref{eq:depinning}) for elastic
interfaces driven with a force $f$. A crucial difference between
yielding and depinning concerns the far-field kernel, which is of
Eshelby type and not only features long range decay, but also%
\footnote{Monotonic, or `abelian' interactions, also lead to {\em no-passing} theorems with fascinating implications~\cite{middleton1992a,SethnaDKKRS93}.}
is non-monotonic~\cite{Baret02,Picard2004}. For example in 2D the elastic kernel can be written as
 \begin{equation}
\label{ker}
     f_{el}[\epsilon(r,t)] \sim \int d r \frac{\cos 4 \phi}{r^2} \delta \epsilon
 \end{equation}
Here, $r$ is the distance from the shear transformation, $\phi$ the angle associated to the position $\vec r$, and $f_{el}$ is the component of the stress tensor projected on the direction of the applied shear.%
\footnote{A more realistic description of plastic interactions is tensorial~\cite{budrikis2017universal}, yet the scalar approximation described in the main text does not affect critical properties near the transition.}
The presence of the positive and negative interactions from the cosine term is what makes $f_{el}$ non-monotonic~\cite{middleton1992a}
-- the transition increases the stresses
for some neighbors, and decreases the stress for others. Indeed, after a
plastic instability, the stress redistribution is on average zero: some
regions see their stress increase, while in others the stress
diminishes. This effect does not occur for the pandemic model or for the
depinning transition, where the motion of a region of the interface can
only destabilize other regions. For a yielding amorphous material, the
stress that a given spot feels in time rises and falls, so that survival
without going unstable becomes increasingly unlikely for the
particularly weak spots. In section~\ref{section:stability} we will see
that the resulting {\em pseudogap} in the probability of regions ready
to yield gives scale-invariant avalanches in the entire yielding solid
phase -- the system shows generic scale invariance.

\Subsubsection{Elasto-plastic models of depinning and yielding}
\label{subsec:BlockModels}

Models of depinning and plastic yielding (sections~\ref{subsec:Depinning}
and~\ref{subsec:Plasticity}) often involve coupled networks of sites
that each slip or jump to new configurations when pushed by one another
or an overall stress field. Coupled block-spring models were introduced in the
study of earthquakes by Carlson and Langer~\cite{CarlsonL89}
without explicit disorder. We can model a depinning system with a $D$-dimensional
lattice of blocks connected by springs, moving in a vertical direction $u$
perpendicular to the lattice. We add a force threshold
$\sigma^y_i = -f_\text{dis}(u_i,r_i)$ to each block that depends
randomly on its vertical position, and have each block move by one unit, $u_i \to u_i + 1$,
 when it goes unstable. This produces a kick that can destabilize
one or more of its neighbors, triggering an avalanche. Adding longer range
springs with strength $G_{ij} \sim r_i-r_j)^{-(1+\alpha)}$ gives us a long-range
depinning model, and adding springs $G_{ij} = \cos(4 \phi)/(r_i-r_j)^2$
gives us a model of yielding in amorphous materials\cite{Baret02,Picard2004,nicolas2018deformation,cao2018a}. We shall return to this
class of models again when we discuss mean-field theories of
depinning in sections~\ref{subsubsec:DepinningMeanField}
and~\ref{subsubsec:YieldingMeanField}.

\Subsubsection{Jamming transition}

\begin{figure}[t]
\centerline{\includegraphics[width=12cm]{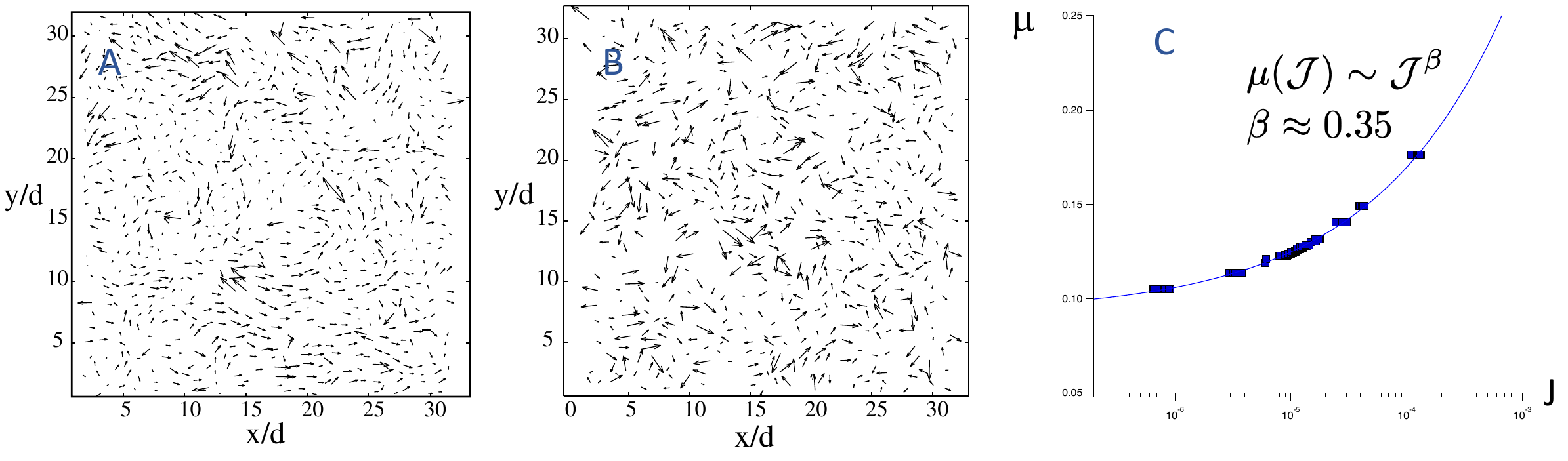}}
\caption{{\bf Jamming:} Critical properties of experimental granular flows. Fluctuating velocity of particles at the free surface of a granular flow in an inclined plane geometry of angle
(a)~$21^o$ degree close to $\Theta_\text{c}=20^o$, showing long-range spatial correlations and (b)~far away from it $\Theta= 26^o$, where the flow is faster, less dense, and where these correlations are absent. From~[\citen{Pouliquen04}]. (c)~Macroscopic friction coefficient $\mu$ versus dimensionless shear rate ${\cal J}$ for different layer thicknesses $h$, showing an exponent $\beta\approx 0.35$. This experiment is performed is a drum geometry with short-range electrostatic repulsion between suspensed particles, to ensure that they are effectively frictionless. We shall study the singular flow rates produced by the avalanches for depinning, yielding, and jamming in Secs.~\ref{subsec:DepinningTheory}, \ref{subsubsec:YieldingScaling}, and~\ref{subsubsec:JammingScaling}. From~[\citen{perrin2021nonlocal}].}
\label{fig:grain}
\end{figure}

How can a crowded, dense collection of particles manage to move and avoid each other as they are sheared?
This question is relevant for the glass transition (when a liquid becomes an amorphous solid), the study of the flow of pedestrians~\cite{PedestrianJamming} or (unexpectedly) the optimization of neural networks~\cite{WyartNeuralNetworkJamming}. Initially it was motivated by granular materials, which can be solid or liquid depending on density and shear stress. They present a yield stress $\Sigma_\text{c}$ separating these two phases, which for cohesionless materials must be proportional to the pressure $p$ (the only stress scale in the problem)~\cite{Andreotti13}. From the ratio $\mu_\text{c}=\Sigma_\text{c}/p$, one obtains the angle of repose $\Theta_\text{c} = \arctan{\mu_\text{c}}$, at which a layer of sand stops flowing. The stress ratio $\mu_\text{c}$ and the
resulting angle of repose%
\footnote{As mentioned above,  sandpiles of frictional particles present two characteristic angles: $\Theta_\text{c}$ where flow stops, and $\Theta_\text{start}>\Theta_\text{c}$  where flow starts \cite{NagelAngleOfRepose,nowak2005maximum,perrin2019interparticle,Dijksman11, mowlavi2021interplay,DeGiuli17a,Zapperi22}. }
$\Theta_\text{c}$ determines a critical point.
For angles just above $\Theta_\text{c}$, flow can occur despite the material being almost as dense as in the solid phase. Yet to avoid each other, the motion of the particles must be very cooperative~\cite{Pouliquen04,Olsson10,During14} with
long-range correlations or `eddies', as illustrated in Fig.~\ref{fig:grain}A.
Also, the flow curve is singular at $\mu_\text{c}$~\cite{Andreotti13,Olsson11,degiuli14d,perrin2021nonlocal} as illustrated in Fig.~\ref{fig:grain}B. For example in dense suspensions $\mu({\cal J})-\mu_\text{c}\sim {\cal J}^\beta$, where ${\cal J}\equiv \eta_0 \dot\epsilon/p$ is a dimensionless ratio of the shear rate $\dot\epsilon$, the viscosity $\eta_0$ of the suspending fluid, and the particle pressure $p$
(see section~\ref{subsubsec:YieldingScaling}).

\begin{figure}[t]
\centerline{\includegraphics[width=12cm]{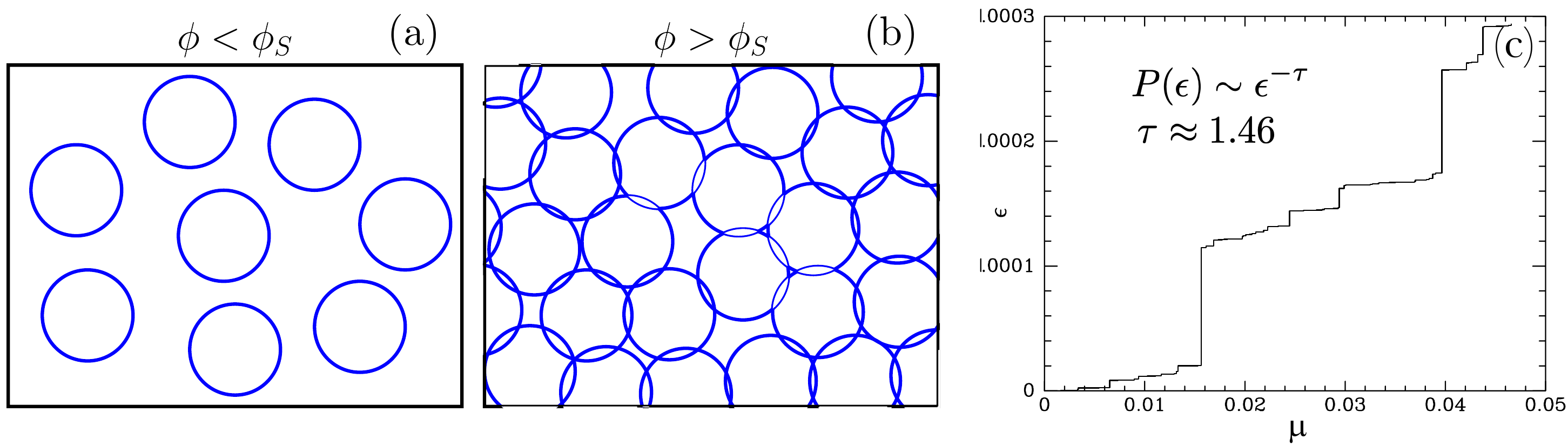}}
\caption{
(a)~Unjammed assembly of particles at low packing fraction
$\phi<\phi_\text{c}$ (b)~Jammed packing with $\phi>\phi_\text{c}$.
(c)~At $\phi_\text{c}$, as the stress anisotropy $\mu$ is increased, the
strain $\epsilon$ displays a devil staircase: there are sequence where
no rearrangements occur (horizontal line) followed by sudden jumps in
strain or `avalanches' that are power-law distributed. We will associate
the presence of such a crackling phase with its marginal stability. From
Ref.~\cite{Combe00}.
}
\label{fig:jamming}
\end{figure}

Progress was made by considering an ideal system: frictionless particles
with repulsive, finite-range
interactions (e.g., slippery rubber balls)~\cite{Liu10,Hecke10}. Imagine their positions to be initially random, then moved to locally minimize the energy. As illustrated in Fig.~\ref{fig:jamming}, for a large enough initial packing fraction of particles, this procedure leads to an amorphous solid, similar to a structural glass. For a small density, it leads to a gas of particles. At some threshold value $\phi_\text{c}$, the particles are barely touching (a good model of granular materials)~\cite{OHern03}.
Key findings at $\phi_\text{c}$ are that (i)~there exists a $\mu_\text{c}\approx 0.05$ below which the material is solid~\cite{Peyneau08,Lespiat11}. For $\mu>\mu_\text{c}$, flow occurs and the flow curve is singular, capturing quantitatively the experimental results of Fig.~\ref{fig:grain}C. (ii)~Avalanches are observed for $\mu<\mu_\text{c}$, as shown in Fig.~\ref{fig:jamming}C~\cite{Combe00}. (iii)~Some structural properties display singular behavior. In particular, the distribution $P(f)$ of contact forces $f$, or the distribution $g(h)$ of interstices between particles, are characterized by non-trivial exponents at small arguments~\cite{Lerner13a}:
     \begin{eqnarray}
     P(f)&\sim& f^\theta \ \ \hbox{with} \ \theta\approx 0.44 \ , \\
    g(h)&\sim& h^{-\gamma} \ \ \hbox{with} \ \gamma\approx 0.38 \ .
     \end{eqnarray}

\Subsection{Instability thresholds and stability}
\label{section:stability}

Avalanches are triggered at weak spots, where an abrupt nonlinear response
is triggered by an increasing field. More generally, amorphous and disordered
materials allow for low-energy local rearrangements that can have a
distribution $P(z)$  of excitations whose threshold force is $z$. It has long been known that
systems like Coulomb glasses~\cite{Efros75} and 
spin glasses~\cite{Thouless77} will excite those rearrangements with the
lowest energies to optimize the interaction energies in finding the 
ground state. This removes many of the lowest energy excitations, leaving 
a density $P(z)$ of excitations that vanishes as a power law,
\begin{equation}
P(z)\sim z^{\theta_y} \ ,
\label{eq:pseudo}
\end{equation}
as the local excitation energy $z$ goes to zero. 
Such a vanishing density is called a {\it pseudo-gap} (to distinguish it from a hard gap with no excitations in a finite energy range). Since then, it was realized that this situation is common, and affects both avalanches and  stationary flows when they occur for large forcing.

\Subsubsection{Nature of excitations in the systems considered}
Excitations correspond to nearly unstable rearrangements of the material. In magnetic systems, they are  spins that are weakly polarized \cite{Thouless77}.
In elastic manifolds, they correspond to  small portions of the interface. 
Here we focus on systems with long-range interactions, in which the notion of excitations  turns out to be more crucial.

{\bf  Amorphous solids:}
 As discussed above, instabilities in that case are shear transformations, illustrated in Fig.~\ref{fig:excitations}(b). 
 Here $z$ is the increment of shear stress that will trigger an
instability of a shear transformation zone, and $P(z)$ is the density of zones
with trigger stress $z$. As briefly mentioned in
Sec.~\ref{subsec:Plasticity}, the non-monotonicity in the long-range interactions
between transformations leads to a local stress history that varies both
up and down as the material is plastically deformed, depleting 
the density of the lowest threshold transformation zones. As in the 
equilibrium systems, sheared amorphous solids are
found \cite{Karmakar10a,Lin14a} to displays a pseudo-gap, with 
$P(z)\sim z^{\theta_y}$.
Interestingly, $\theta_y$ varies continuously under loading \cite{Lin15a}. The condition $\theta_y>0$ can be obtained form a stability argument \cite{Lin14a}. Assume that $\theta_y=0$ such that $P(x=0)=P_0>0$.  Consider a shear transformation occurring at the origin,  leading to a kick of stress of order $\Delta \Sigma(r)\sim r^{-d}$ at a distance $r$. The probability that a region of unit area displays an instability in response to this kick is $\sim \Delta \Sigma(r) P_0$. In a system of size $L$, the total number of new instabilities $R_0$ is then obtained by integration: $R_0\sim P_0\int_{r<L} \Delta \Sigma(r) r^{d-1}dr\sim \ln(L)$. This result implies that an individual plastic events always completely destroy the system (since $R_0\gg 1$), with probability one. We know from experience that amorphous solids are stable to such local perturbations, thus implying $\theta_y>0$.

{\bf Packings of frictionless hard spheres:} These systems are {\it isostatic}: they just have enough contacts to ensure the stability of the material~\cite{kane2014topological}.
Under loading undeformable particles, rearrangements can only occur when a contact force goes to zero, leading to the opening of a contact and to a global motion of all particles called a `floppy mode'. An  example of such an excitation is shown in Fig.~\ref{fig:excitations}(b). This motion stops when a new contact is formed. This process gives a kick of stress in the entire system, which can in turn trigger the opening of new contacts. The magnitude of the kick can be shown to be proportional to the displacement along the floppy mode before a new contact is formed, which occurs when a small gap between nearly touching particles closes. Thus, this distance is controlled by the distribution $g(h)\sim h^\gamma$ of gaps.  Furthermore, the probability that a  kick of given magnitude opens a new contact  is controlled by the distribution of small forces $P(f)\sim f^\theta$. Requiring stability, i.e. that the number of new excitations triggered by a single one does not diverge in the thermodynamic limit (i.e. $R_0\leq 1$), can be shown to imply \cite{Wyart12}: 
\begin{equation}
    \gamma\geq 1/(2+\theta) \ .
    \label{eq:margin}
\end{equation}
As described below, infinite range interactions (independent of distance) imply \cite{Muller14} a saturation of Eq.~\eqref{eq:margin}, $\gamma = 1/(2+\theta)$, in agreement with numerical estimates of $\gamma$ and $\theta$. A great success of replica theory applied to the jamming transition is the computation of $\theta\approx 0.42$ and $\gamma\equiv 1/(2+\theta) \approx 0.41$ in infinite dimensions \cite{Charbonneau14}. These exponents agree well with simulations in finite dimensions~\cite{Lerner13a,Charbonneau15}, as perhaps expected from numerical evidence~\cite{GoodrichLiuUCD2,SartorRCnn} that the upper critical dimension is $D_\text{uc} = 2$ for frictionless jamming.

\Subsubsection{ Pseudo-gaps imply crackling noise}
\label{subsec:PseudoGapCrackling}

Stability arguments do not exclude the possibility that the density of excitations presents a hard gap (and strictly no rearrangements for a finite range of loading).  Even if gapped configurations  exist, they are not found in practice.  Fig.~\ref{fig:excitations}(a) gives a pictorial illustration as to why it is so: for some systems and dynamics the system essentially stops as soon as it becomes stable, leading to a gapless marginal state. In mean-field spin glasses for example, it can be shown that finding gapped configurations is a NP complete problem~\cite{behrens2022dis}. 

The existence of a pseudo-gap implies that the rearrangements must be collective -- avalanche like, with an initial rearrangement triggering at least some large events, leading usually to what we call crackling noise, without tuning or self-organizing to a critical point.
Following~[\citen{Muller14}], consider for example an amorphous solid of volume $N$, and assume that Eq.~\eqref{eq:pseudo} applies. During an adiabatic loading, there will be periods without plasticity: their magnitude $\delta \Sigma$ is given by $z_{min}$, the stability of the least stable excitation. Standard extreme value statistics arguments \cite{Karmakar10a,Lemaitre07a} lead to the typical value 
$z_{min}\sim N^{-1/(1+\theta_y)}\gg \frac{1}{N}$.  It implies that the rate of plastic events is sub-extensive: a system twice as large does not display twice as many plastic events. Assume that no crackling occurs: then for local excitations, a plastic event involves ${\cal O}(1)$ excitations and thus dissipates an energy ${\cal O}(1)$, which is independent of $N$. For a global change of stress $\Delta \Sigma$, the energy dissipated is thus $\sim \Delta \Sigma/\delta \Sigma\sim N^{1/(1+\theta_y)}\ll N$, {i.e.} it is sub-extensive. Thus, in the absence of crackling, dissipation per unit volume must vanish in the thermodynamic limit. 
\mw{Plastic materials displaying a finite density of dissipation (which we expect to be the generic situation for disordered materials) must then display crackling. In particular}
for dissipation to be extensive, one must have $\langle S\rangle \sim N^{\theta_y/(1+\theta_y)}$.

We return to infinite range interactions (such as frictionless hard spheres in any dimensions, or the SK model of spin glasses). In that case, it can be shown that if inequalities like Eq.~\eqref{eq:margin} are not saturated~\cite{Muller14}, then an excitation has a vanishing probability to trigger a second one in the thermodynamic limit: crackling cannot occur.  But we have seen before that crackling noise must occur (to allow for an extensive dissipation), thus our hypothesis is incorrect: we conclude that the bound in Eq.~\eqref{eq:margin}  must be saturated when infinite-range interactions are present. 
Note that saturation does not occur for power-law interactions, e.g. in amorphous solids.

Ultimately, this approach leads to a classification of glassy systems based on the range of the interaction of their excitations. If sufficiently long-range and \mw{non-monotonic}, then a pseudo-gap must be present, and crackling occurs. Instead for short-range interactions, crackling occurs only at a critical point and not in the entire glassy phase. As we shall see below, the pseudo-gap exponent also characterizes the flowing phase, when it exists.

\Subsection{Theoretical Results: Depinning}
\label{subsec:DepinningTheory}

\Subsubsection{Mean-field theory of depinning}
\label{subsubsec:DepinningMeanField}

Depinning (see Eq.~\eqref{eq:depinning}) is characterized by the competition between a decaying elastic force, \ar{$f_{\text{el}} \sim \int d^Dr  (u(r')-u(r))/|r-r'|^{D+\alpha}$} that flattens the interface and a disorder force of the $D+1$ dimensional medium that tries to roughen it.  Dimensional analysis can  help to determine the result of this competition, if we assume that  the interface is self-affine, namely that after a dilatation $x \to b x$, its transverse fluctuations  grow as  $b^{\zeta}u(x)$. After  the dilatation, the elastic force is rescaled as $b^{\zeta - \alpha}$ for $\alpha < 2$ (for shorter-ranged forces it turns out to scale as $b^{\zeta - 2}$) and the  short-range  disorder   force  turns out to be rescaled  as $b^{D/2-\zeta/2}$.  Using a Flory type of argument one has to balance the two  forces  in order to estimate the roughness exponent \cite{nattermann1990,agoritsas2016driven}. Thus, the interface results flat ($\zeta = 0$) at the upper critical dimension $D_\text{uc}=2\alpha$ for $\alpha<2$, matching nicely with the value $D_\text{uc}=4$ above $\alpha>2$.
 
Above $D_\text{uc}$ disorder is not able to roughen the interface, which turns out to slide with a velocity linear in the force above threshold ($\beta=1$), and display mean field avalanches ($\tau=3/2$, $\sigma= 1/2$, and $z=\alpha$), as discussed below. Below $D_\text{uc}=2\alpha$, the interfaces are rough ($\zeta>0$), the velocity grows with a power $\beta$ less than one, with avalanche exponents $\tau$, $\sigma$, and $z$ smaller than those in mean-field.

This Flory argument correctly predicts  $D_\text{uc}$, a result that  can  be also obtained treating the disorder as a perturbation \cite{larkin1970,tanguy1998,chauve2000,cao2018}. Unfortunately both arguments fail in predicting the exact value of the exponents below $D_\text{uc}$. Indeed, for purely linear models, the Flory balance works perfectly at each length scale. However, any nonlinear terms will couple the different length scales, and thus demand a renormalization group approach.%
 \footnote{In systems with `KPZ'-type nonlinearities Flory's arguments do not work in any dimension. In this case
 it remains controversial if it exists an upper critical dimension above which interfaces are flat.}

A mean-field theory for depinning can be formed using the block-spring
model of section~\ref{subsec:BlockModels}, but having every pair of sites
$x_i$ and $x_j$ connected by a spring $G_{ij} \equiv G/N$, with $N$ the 
number of blocks. After each block slips, all blocks receive a kick (force
increment) of $G/N$.
Hence, the dynamics are precisely the same as that in the 
Bienaym\'e-Galton-Watson process we saw in the pandemic model. The value
of the force $f$ controls the parameter $R_0$, which becomes critical ($R_0=1$) at $f=f_c$. Because all of the kicks are in the same direction, the \mw{interaction is monotonic}, and there is no pseudogap. 
Above $f_c$ a stationary flow is present with a permanent fraction of unstable blocks $\sim |f-f_c|^\beta$, with $\beta=1$.

\Subsubsection{Finite dimension and Functional RG}
\label{subsubsec:FunctionalRG}

\begin{table}[t]
\tbl{{\bf Critical exponents and scaling relations} for depinning systems of $D$
dimensional interfaces in $D+1$ dimensions, assuming no nonlinearities
of the qKPZ type and \ar{elastic forces $f_{\text{el}} \sim \int d^Dr  (u(r')-u(r))/|r-r'|^{D+\alpha}$}. The last column shows one-loop functional
renormalization group
calculations~\cite{narayan1992,nattermann1992,narayan1993,ertas1994}
with $D_\text{uc} = \min(2\alpha, 4)$ \ar{and $\epsilon= D_\text{uc}-D$}.
Above $D_\text{uc}$ behavior is mean-field. For  $D<D_{\text{uc}}$,
exponents are measured numerically with a precision at the second digit.
Note that depinning models do not have a  pseudogap, so $\theta_y=0$.
}
{\begin{tabular}{cccccc}
  \hline
  \hline
  \textbf{Depinning}  &  \textbf{Observable} &  $D=1$   &  $D=1$   & $D=2$  & \textit{FRG}  \\
  \textbf{exponent}  & & $\alpha=2$ & $\alpha=1$ & $\alpha=2$  &  $ D = D_\text{uc} -\epsilon$  \\
  \hline
  $z$ & $ t(L) \sim L^z$ & 1.43 & 0.77 & 1.56  & $D_\text{uc}/2-2 \epsilon/9$ \\
  $\zeta$ & $u(x) \sim x^{\zeta}$ & 1.25 & 0.39 & 0.75 & $\epsilon/3$ \\
  $\tau$ & $P(S) \sim S^{-\tau}$ & \multicolumn{3}{c}{$\tau=2-\alpha/(D+\zeta)$} & $ 3/2 -\epsilon/(3 D_\text{uc})$ \\
  $\nu$ & $\xi \sim |f-f_c|^{-\nu}$ &   \multicolumn{3}{c}{$\nu=1/(\alpha-\zeta)$}  & 
  $2/D_\text{uc}+4\epsilon/(3 D_\text{uc}^2)$ \\
  $\beta$ & $v \sim |f-f_c|^{\beta}$ & \multicolumn{3}{c}{$\beta=\nu(z-\zeta)$}  & $1- \epsilon/(9 D_\text{uc}) $ \\
  \hline
  \hline
  \end{tabular}}
\label{tab:dep}
\end{table}

Table~\ref{tab:dep} gives a summary of exponents for the depinning of
$D$-dimensional manifolds in $D+1$ dimensions. The last three rows in
the table give a set of scaling relations that allow one to express all
the exponents in terms of two, $\zeta$ and $z$, valid below
$D_\text{uc}$. For example, the velocity above threshold can be
expressed as the ratio between avalanche displacement and avalanche
duration, namely $v\sim  \xi^\zeta/\xi^z \sim |f-f_c|^{-\nu(\zeta-z)}$,
giving the equation for $\beta$. The last column in Table~\ref{tab:dep}
gives one-loop functional RG (FRG) results for the
exponents~\cite{narayan1992,nattermann1992,narayan1993,ertas1994} close
to the upper critical dimension. Two-loop
\cite{chauve2001,ledoussal2002,ledoussal2004} and three-loop
calculations for the static problem are now
available~\cite{wiese2018field,husemann2018field}.

The RG  approach  captures the large-scale scale invariant properties of a population of configurations. 
The traditional RG studies the flow in a space of parameters (field $H$, temperature $T$, \dots) which yield an effective coarse-grained free energy; the free energy encodes the Gibbs weight of different configurations. Here, the weight of the dynamical trajectories is encoded into the Martin-Siggia-Rose action, which averages trajectories over disorder. The resulting RG involves a full function: the force-force disorder correlator, $\Delta(u-u')$, where $u,u'$ are two different positions of the center of mass of the interface. The fixed point $\Delta^*$ can be computed by FRG as a  perturbation  in  $\epsilon= D_{\text{uc}}-D$ and has a peculiar cusp at the origin $u=u'$ which has been experimentally measured \cite{wiese2021force}.
Indeed, when an avalanche occurs, much of the interface is left pinned at the same position (same disorder), but the disorder force changes abruptly in the regions which have slipped forward (Fig.~\ref{fig:Depinning}), leading to the cusp.

For physical systems in $D=1$ and $D=2$, the best estimation of depinning exponents come from numerical calculations \cite{leschhorn1997,tanguy1998,rosso2002,rosso2003,duemmer2005,duemmer2007,ferrero2013}, summarized in the first two rows of Table~\ref{tab:dep}. The two-loop expansion \cite{chauve2001,ledoussal2002,ledoussal2004,rosso2007}, however, settled a {\em qualitative} question --- showing that the interface at thermal equilibrium has a different roughness and universality class from that of the interface at depinning. Even more surprisingly, FRG \cite{chauve2000} is also able  to describe the thermally activated slow dynamics of magnetic domain walls  driven by a magnetic field much smaller than the critical force. Most of the predictions for this creep regime have been tested both experimentally \cite{lemerle1998,caballero2018} and numerically \cite{ferrero2020}. \jps{See also the FRG results for the avalanche shape in section~\ref{subsubsec:AvalancheTemporalShape}.}

Critical exponents are neither  the only quantity of interest nor the simplest to observe in experiments \cite{laurson2009,bares2019}. The value of the critical force $f_c$  and its universal fluctuations \cite{bolech2004,demery2014,fedorenko2006} are important for applications. In the following we will see in some details some other important example of such observables.

\Subsubsection{The average avalanche time series}
\label{subsubsec:AvalancheTemporalShape}

\begin{figure}[t]
\centerline{\includegraphics[width=12cm]{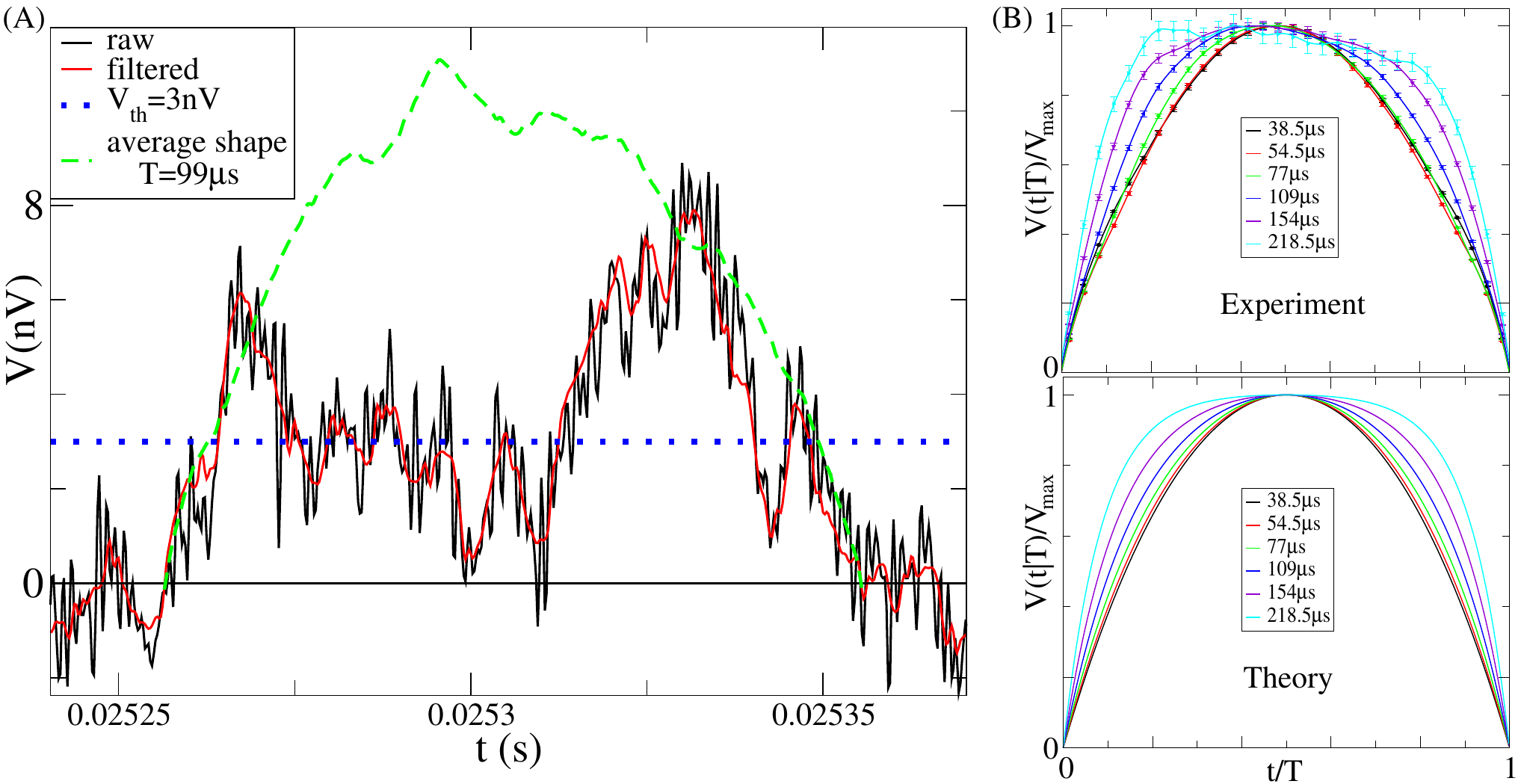}}
\caption{{\bf Avalanche, and average temporal shape,} 
from Ref.~[\citen{PapanikolaouBSDZS11}].
(A)~Experimental signal
during an avalanche (jagged), with noise filtered out (wavy), rather than 
using thresholding (dotted). The dashed
line shows the signal $V(t|T)$ averaged over all avalanches of similar duration.
(B)~Average temporal shape in experiment and predicted by mean-field theory
(Eq.~\eqref{eq:TemporalShapeTheory}) for various avalanche durations.
}
\label{fig:TemporalShape}
\end{figure}

A longstanding challenge for the scaling theory of dynamical disordered systems was the rather symmetrical predictions for the average temporal shape~\cite{KuntzS00,MehtaMDS02}, which disagreed with a much larger asymmetry in experiments~\cite{SethnaDM01}. This eventually was explained as an effect of eddy currents~\cite{ZapperiCCD05}.
Figure~\ref{fig:TemporalShape}A shows the time-dependent mean front velocity during a mean-field magnetic avalanche in a thin magnetic
film~\cite{PapanikolaouBSDZS11}, avoiding the effects of eddy currents.
Note the characteristic
irregular, fractal shape -- almost stopping several times. (If it had stopped,
it would form a smaller avalanche -- big avalanches are just smaller
avalanches that did not stop a few times.) Note also the dashed curve -- an
average $V(t|T)$ over all the experimental avalanches whose durations $T$
were similar to this one. The universal scaling prediction for this average shape is~\cite{KuntzS00,SethnaDM01}
\begin{equation}
\label{eq:TemporalShapeTheory}
V(t|T,r) = T^{d_f/z-1} \curlyV(t/T, T r^{z \nu})
  = T \curlyV(t/T, T r) \ .
\end{equation}

Figure~\ref{fig:TemporalShape}B shows a scaling collapse of average avalanche shapes for
different durations (top) and the universal scaling prediction from
mean-field~\cite{PapanikolaouBSDZS11,LeDoussalWieseAverageTemporalShape}
(bottom). 
The average shape is an inverted parabola for avalanches small compared to the
cutoff $T_\mathrm{max} \sim r^{-z \nu} = 1/r$, and has an explicit analytical
form for longer durations in mean-field theory (Fig.~\ref{fig:TemporalShape}B).

The same functional RG methods we discussed in section~\ref{subsubsec:FunctionalRG}, remarkably, have been applied~\cite{Dobrinevski15avalanche} to predict the average temporal shape both at fixed duration and
fixed avalanche size for $D\leq D_{uc}$.
They predict a slight skewing of the avalanches that nicely fits \ar{the measured shapes}~\cite{MehtaMDS02,durin2016quantitative}, \jps{and also incorporate eddy currents into the scaling predictions~\cite{dobrinevski2013statistics,durin2016quantitative}.}

\Subsubsection{Clusters in avalanches with long-range interactions} 

As we saw in Fig.~\ref{fig:Depinning}, avalanches in systems with
long-range interactions can be disconnected into {\em clusters}. It is
infeasible in most experiments to visualize avalanches one-by
one\cite{lepriol2020}, and naturally impossible to figure out after
several avalanches which clusters came from which avalanches -- we
cannot directly find $\tau$, $\sigma$, etc.

At moderate rates of strain, one can measure the statistics of clusters
localized in space, which fortunately also display universal scaling
\cite{maloy2006}. For example their size $S_c$ has a power law
probability distribution $P(S_c) \sim S_c^{-\tau_c}$. The other
avalanche exponents can be defined for clusters as well. A solid
numerical conjecture predicts the value of cluster exponents in terms of
the avalanche exponents \cite{laurson2010}. For example the relation
$\tau_c= 2 \tau-1$ should hold for $\alpha<2$. One must note, though,
that our analytical study of a simple epidemic model with  long range
dispersal  does not show this behavior~\cite{cao2022clusters}.

 The key to this conjecture is the identification of a
Bienaym{\'e}-Galton-Watson process describing the statistics of the
cluster number \cite{lepriol2021}.  This process has been clearly
observed  numerically for all $\alpha<2$, but no proof of its existence
yet exists. For $\alpha > 2$ the number of clusters is strongly
suppressed; $\tau_c=\tau$ for forces of finite range. It is not yet
clear how $\tau_c$ behaves for large, but finite values of $\alpha$. A
future challenge is to understand the origin of the
Bienaym{\'e}-Galton-Watson process describing the cluster statistics.

\Subsection{Theoretical Results: Yielding transition}
\label{subsec:YieldingTheory}

\Subsubsection{Mean field models of yielding}
\label{subsubsec:YieldingMeanField}

The mean-field version of the elasto-plastic block-spring models for the yielding transition (section~\ref{subsec:BlockModels}) have non-monotonic interaction matrices $G_{ij}$ with both positive and negative values that are randomly chosen with the same distribution and independent of $i$ and $j$ (infinite range).  Let us call $z_i = \sigma_i^y-\sigma_i$, where $\sigma_i$ and $\sigma^y_i$ are respectively the shear stress and the yield stress of that block. When $z_i<0$ the block yields, reducing all the other $z_j$ by $G_{ij}$

In the {\it Hebraud-Lequeux} model \cite{Hebraud98}, $G_{ij}$ is assumed to follow a Gaussian distribution of variance $\sim N^{-1/2}$, where $N$ is the number of blocks. As a result, each variable $z_i$ is a sum of Gaussian random kicks, where the walk ends when the sum hits zero -- a  random walk with an absorbing condition at $z=0$. After yielding, the site is reinserted at some finite positive $z_i$ value. Solving this random walk problem, we find that the distribution $P(z)$ evolves until it vanishes linearly with $\theta_y=1$.  The flow curve is found to be singular at $\Sigma_c$, with $\beta=2$. Interestingly \cite{jagla2015avalanche}, the distribution of avalanches sizes depends on details of how the system is loaded. For realistic loading, $\tau\approx 1$. 

In a more realistic mean-field model, the noise is not assumed to be Gaussian. It is still  i.i.d. and thus has no spatial correlations, but the distribution of $G_{ij}$ is chosen to satisfy that of the true propagator. For a propagator that vanishes with distance as a power-law, $P(G)$ is then power-law distributed \cite{Lemaitre07}. In that case, $z_i$ follows a {\em Levy flight} with a Levy coefficient $\mu=1$. The pseudo-gap is still governed by an absorbing condition: it is found that after a quench at zero stress, $\theta_y=1/2$  \cite{Lin16}. However, here $\theta_y$ continues to evolve as the system is strained; it starts by rapidly decreasing, and then increases. This has been observed in finite dimensional elasto-plastic models \cite{Lin15a} and in molecular dynamics of amorphous solids \cite{ji2019theory,ozawa2018random}. Flow curves are again singular, with $\beta=1$ \cite{lin2018microscopic}. In this model, aging responses following a quench (a minimization of the energy from some random configuration at $t=0$) can be computed \cite{parley2022mean}, and are found to be similar to experimental ones.

\Subsubsection{Scaling theory for yielding}\
\label{subsubsec:YieldingScaling}

In this section we will provide a link between the avalanche physics
at the mesoscale and the flow curve at the macroscale, for systems 
with non-monotonic long-range interactions (such as amorphous materials
yielding under stress, section~\ref{section:stability}).

{\it Stationarity:} Consider applying a quasi-static strain in a system
of linear size $L$ at the yielding transition. The stress $\Sigma$
ramps up under increasing strain, and drops abruptly during avalanches. 
On average,
these must cancel out \cite{Maloney04,Karmakar10a,Salerno13}. An avalanche of size $S$ will lead to a stress drop in the system that is proportional to 
$\delta \Sigma \sim \langle S \rangle/L^d$. The
mean avalanche size is
$\langle S \rangle\sim \int^{S_\text{max}} S\times S^{-\tau} dS \propto S_\text{max}^{2-\tau}$, where $S_\text{max}\sim L^{d_f}$ is the
largest avalanche that can fit into the system. The stress rise between
avalanches is determined by the pseudogap in the distribution of threshold
stresses. As discussed in section~\ref{section:stability}, the 
weakest site will yield after the stress increases by 
$\sim L^{-d/(1+\theta_y)}$. Equating the average drop
and the average increment, one obtains\cite{Lin14}:
\begin{equation}
\label{s1}
\tau=2-\frac{\theta_y}{\theta_y+1}\frac{d}{d_f} \ .
\end{equation}

{\it Dynamics in the flowing regime:}
As one approaches the critical stress $\Sigma_c$ from above in the flowing
regime, there is a diverging length scale $\xi\sim (\Sigma-\Sigma_c)^\nu$.
Below this length scale, both in depinning transitions~\cite{Fisher98}
and in the yielding transition~\cite{Lin14} one can roughly describe the 
motion as a continuously evolving collection of avalanches of spatial
length $\xi$, covering space.
The net strain rate $\mathcal{\dot\epsilon}$ is the
number of avalanches $\sim 1/\xi^d$ times the strain release $S\sim \xi^{d_f}$
per avalanche divided by the duration $T \sim \xi^z$ of these avalanches,
hence $\dot{\epsilon}= S/(T\xi^d) \sim(\Sigma-\Sigma_c)^{\nu(d-d_f+z)}$.
Hence the strain rate $\dot{\epsilon}\sim (\Sigma - \Sigma_c)^\beta$ grows
with an exponent
\begin{equation}
\label{eq:beta}
\beta=\nu(d-d_f+z) \ .
\end{equation}

{\it Stress fluctuations:} Continuing to view the flow as a superposition of avalanches of size $S_\text{max}$ spanning boxes of size $V=\xi^d$, each will lower the stress in its box by an amount $\delta \Sigma\sim S_\text{max}/\xi^d \sim (\Sigma-\Sigma_c)^{\nu(d-d_f)}$. Making the common assumption that all quantities with the same units share the same scale near a critical point, we expect 
that the fluctuations of stress $\delta \Sigma$ on the scale $\xi$ must be of order of the distance to threshold $\Sigma-\Sigma_c$, leading to \cite{Salerno13,Lin14}:
\begin{equation}
\label{s3}
\nu=1/(d-d_f) \ .
\end{equation}
Putting Eqs.~\eqref{s3} and \eqref{eq:beta} together, one obtains
$\beta=1+z/(d-d_f)\geq 1$. Note that the scaling relations in
Eqs.~\eqref{s1}--\eqref{s3} should work  in elasto-plastic models,
particle based models, and real systems, as supported by observations
\cite{Lin14,nicolas2018deformation}. One must note that elasto-plastic
models assume that elastic interactions propagate instantly, and
sometimes give an unphysical value of $z<1$ (leading to avalanche
propagation faster than the speed of sound). In this case, we presume
that $z=1$ and the value of $\beta$ in Eq.~\eqref{eq:beta} should thus
be larger in physical systems than in elasto-plastic
models~\cite{lin2018microscopic}.

These scaling arguments can be extended to non-stationary situations \cite{popovic2021scaling} such as the slow creep flows \cite{balmforth2014yielding} that follows a sudden increase of stress, to non-local phenomena \cite{goyon2010does} where $\xi\sim |\Sigma-\Sigma_c|^{-\nu}$ is found to characterize the length scale on which an obstacle or interface perturbs flow \cite{gueudre2017scaling}, and to finite temperatures \cite{ popovic2021thermally,ferrero2021yielding} that cause a ``thermal rounding'' of the flow curves, which  becomes smooth near $\Sigma_c$.

\Subsection{Theoretical Results: Jamming}
\label{subsec:jammingtheory}

\Subsubsection{Infinite-dimension calculations of avalanches}

Avalanches are dynamical phenomena. Yet `equilibrium avalanches' can be defined, by tilting the energy landscape continuously by adding a force or field, and tracking the position of the global minimum of the system \cite{Le-Doussal12}. It then displays jumps, whose statistics can be computed in infinite dimension using  replica methods. Right at the jamming transition, this method was successfully used to compute $\tau=(3+\theta)/(1+\theta)\approx 1.41 $\cite{franz2017mean} in close agreement with numerical experiments for nonequilibrium avalanches in three dimensions \cite{Combe00}.%
\footnote{It is known in the random-field Ising model that the universality class of the equilibrium system is different from that of the non-equilibrium, avalanche model at zero temperature. This has been shown in dimensions smaller than about 5.1 using remarkable non-perturbative functional renormalization-group calculations~\cite{TarjusNonPerturbativeRG5.1}, and numerically in two dimensions~\cite{HaydenRS19}. Here, however, we are above the upper critical dimension, where  non-equilibrium and equilibrium  properties should be similar.}
 The success of the mean-field approach arguably results from the infinite-range nature of excitations in this system \cite{During14}. 

A system of well-compressed particles (packed more densely than at the jamming threshold) should exhibit a yielding behavior
as described in subsection~\ref{subsec:Plasticity}.
Replica theory of jamming predicts $\tau=1$ 
for the rigid phase~\cite{franz2017mean}. This mean-field result may not hold in three dimensions however, as the length scale of excitations rapidly decreases away from jamming \cite{shimada2018spatial}, and spatial correlations are then presumably important.

\Subsubsection{Suspension and granular flows of frictionless particles}
\label{subsubsec:JammingScaling}

Much of the jamming literature has focused on frictionless particles,
both numerically \cite{Olsson07,Heussinger09,Lerner14,olsson2019dimensionality,Trulsson12} and theoretically \cite{DeGiuli15a}.
Friction is of course important for most of the applications of jamming -- from
soil engineering to pharmaceuticals to geophysics.
It was recently realized that a class of suspensions displaying ``shear
thickening'' display effectively frictionless particles at low stresses, but
they become frictional at large stresses \cite{Mari14,Wyart14,clavaud2017revealing,comtet2017pairwise}. Controlled experimental systems with these properties have been designed \cite{perrin2021nonlocal,perrin2019interparticle,RamaswamyGLSKGSCCnn,RamaswamyGSCCnn}, allowing one to experimentally test scaling 
theories of flow for frictionless jamming.

Consider a frictionless packing at the critical stress anisotropy
$\mu=\Sigma/p=\mu_c$. The first step of the theory consists of
estimating the number of contacts $\delta z$ that open if the stress
anisotropy is increased by $\mu-\mu_c$. One finds, by estimating the
change of forces in the contacts induced by such a perturbation, and
using that the distribution of small force follows $P(f)\sim f^\theta$,
that \cite{DeGiuli15a}:
\begin{equation}
\label{s4}
\delta z\sim(\mu-\mu_c)^{\frac{2+2\theta}{3+\theta} } \ .
\end{equation}
Using this, one can use mechanical considerations to compute how an 
infinitesimal applied strain $\delta \epsilon$ at the boundary moves particles
relatively to each other, by some amount $\delta r$ called the non-affine displacement. Defining ${\cal L}\equiv \delta r/\delta \epsilon$, one finds \cite{DeGiuli15a}:
\begin{equation}
\label{s5}
{\cal L}\sim \delta z^{-\frac{2+\theta}{1+\theta} } \ .
\end{equation}
The overall viscosity of such a suspension is proportional to the dissipated power, which grows as the square of the particle velocities, and is thus proportional to ${\cal L}^2$. The viscous number ${\cal J}$ is inversely proportional to this viscosity, thus ${\cal J}\equiv \eta_0 \dot\epsilon/p\sim {\cal L}^{-2}$. Together with Eqs.~\eqref{s4} and~\eqref{s5},
one obtains the flow curve:
\begin{equation}
\label{s6}
\mu-\mu_c\sim {\cal J}^\frac{3+\theta}{8+4\theta}\approx {\cal J}^{0.35} \ ,
\end{equation}
in  agreement with experiments, as shown in Fig.~\ref{fig:grain}. Several diverging length scales are predicted at the transition \cite{During14}; one of
these (the divergence of non-local effects) has been favorably tested  \cite{perrin2021nonlocal}.

Note that this theory is mean-field in nature, as it neglects spatial correlations that could occur {e.g.} in the structure. Its quantitative agreement with finite dimensional numerical studies \cite{DeGiuli15a} and experiments \cite{perrin2021nonlocal} again underlines the mean-field character of the jamming transition for frictionless particles, where excitations have infinite range interactions at threshold. This situation does not hold for frictional particles, where other exponents are found \cite{Trulsson16,ramaswamy2021universal} and spatial correlations are relevant \cite{DeGiuli17}. An additional complexity is that in the presence of friction, the flow curve is non-monotonic and the transition is first order close to the critical point \cite{DeGiuli17,perrin2019interparticle,Dijksman11, mowlavi2021interplay}.

\Subsection{Conclusion and open questions}
\label{subsec:conclusions}

We have described the closely intertwined physics of avalanches, the stability of the rigid phase, and stationary flows in a variety of disordered systems. Scaling concepts are the key to building such a unifying description. The renormalization group inspires these scaling descriptions, but they are useful even in cases where a RG procedure has not (yet) been developed. We conclude by listing a few open problems in this field:

{\it RG description of the yielding transition:} Although the jamming transition of hard particles appears to be mean-field in character, the yielding transition of amorphous materials appears more subtle. Currently, there is no analytical approach to compute exponents in that case \mw{(except for avalanche exponents in conditions where a narrow shear band appears,  where  a mean-field description may apply~\cite{dahmen2009micromechanical,antonaglia2014bulk,wright2016experimental,wright2018slip}}). 

{\it Avalanches and first order transitions:} Systems often exhibit avalanche precursors before abrupt changes in behavior. 
The abrupt change in behavior is due to a non-monotonic flow curve: the same force (stress) can lead to different velocities (strain rate).
Quasi-brittle materials exhibit power-law fracture precursors under tension~\cite{ShekhawatZS13} before they fracture into two. 
This also occurs in complex fluids (such as loosely connected colloidal gels)
and earthquakes (where frictional forces acting on the fault can decrease with sliding velocity \cite{zheng1998conditions}). 
Inertia can also lead to such flow curves in depinning problems \cite{Fisher1997,Schwarz2001} and in amorphous solids \cite{Salerno12,nicolas2016effects}. Velocity (or strain-rate) weakening can make the transition between a rigid and flowing phase first order and hysteretic, possibly destroying avalanches. Recent work on frictional interfaces suggests that avalanches persist in that situation (although their statistics change), and act as nucleation centers for  system-spanning events \cite{de2022scaling}. In that view, the avalanche size diverges at some threshold stress, beyond which the  rigid material is unstable to the presence of a large nucleus of flowing material. How generic these results are remain to be seen.

{\it Nucleation of failure:} This contribution focuses on stationary
flow in disordered materials. Another important question is how these
materials break as the loading continuously increases.  For amorphous
solids, this usually occurs by forming a shear-band where plastic strain
localizes. Some have argued that the shear band nucleates in a process
similar to that of the random-field Ising model
\cite{ozawa2018random,ozawa2021rare}, or fracture \cite{Popovic2018};
others suggest it is governed by linear stability considerations~\cite{Fielding2021}. Numerical studies~\cite{richard2021finite} report
very slowly-decaying finite-size effects that remain to be understood.

\mw{{\it Yielding at finite temperature $T$ and the glass transition:} we focused on the yielding transition at $T=0$. At finite temperature,   the flow curve loses its singular behavior, in a stress interval around $\Sigma_c$ that vanishes as a power law of $T$ \cite{popovic2021thermally,ferrero2021yielding}. Such  a 'thermal rounding' behavior  can be captured by mean field calculations or scaling arguments \cite{popovic2021thermally}. Yet, a spatial description of thermal avalanches triggered by rare activated events  is still missing. It may be very relevant to understand the collective dynamics that occurs near the glass transition   \cite{chacko2021elastoplasticity}.}


\jps{\section*{Acknowledgment}
JPS would like to acknowledge the support of NSF DMR-1719490.} 
\mw{MW acknowledges support from the Simons Foundation Grant (No.~454953 Matthieu Wyart) and from the
SNSF under Grant No.~200021-165509.}

\bibliographystyle{ws-rv-van}
\bibliography{ch15/ws-rv-sample.bib,Avalanches.bib,Rossobib.bib,SethnaRecs.bib,Wyartbibnew.bib}

\begin{thebibliography}{198}
\providecommand{\natexlab}[1]{#1}
\providecommand{\url}[1]{\texttt{#1}}
\expandafter\ifx\csname urlstyle\endcsname\relax
  \providecommand{\doi}[1]{doi: #1}\else
  \providecommand{\doi}{doi: \begingroup \urlstyle{rm}\Url}\fi

\bibitem{SethnaDM01}
J.~P. Sethna, K.~A. Dahmen, and C.~R. Myers, Crackling noise, \emph{Nature}.
  {\bf 410}, \penalty0 242--250,  (2001).

\bibitem{SethnaDP06}
J.~P. Sethna, K.~A. Dahmen, and O.~Perkovi{\'c}.
\newblock Random-field models of hysteresis,
  \url{http://arxiv.org/abs/cond-mat/0406320}.
\newblock In \emph{The Science of Hysteresis, Vol. II}, pp. 107--179. Academic
  Press,  (2006).

\bibitem{Zapperi22}
S.~Zapperi, \emph{Crackling noise: Statistical physics of avalanche phenomena}.
  (Oxford University Press, Oxford, 2022).

\bibitem{NagelAngleOfRepose}
S.~R. Nagel, Instabilities in a sandpile, \emph{Rev. Mod. Phys.} {\bf 64},
  \penalty0 321--325 (Jan, 1992).
\newblock \doi{10.1103/RevModPhys.64.321}.
\newblock URL \url{https://link.aps.org/doi/10.1103/RevModPhys.64.321}.

\bibitem{Fisher1997}
D.~Fisher, K.~Dahmen, S.~Ramanathan, and Y.~{Ben-Zion}, Statistics of
  {{Earthquakes}} in {{Simple Models}} of {{Heterogeneous Faults}}, \emph{Phys.
  Rev. Lett.} {\bf 78}\penalty0 (25), \penalty0 4885--4888,  (1997).
\newblock \doi{10.1103/PhysRevLett.78.4885}.

\bibitem{DahmenEarthquakes06}
A.~P. Mehta, K.~A. Dahmen, and Y.~Ben-Zion, Universal mean moment rate profiles
  of earthquake ruptures, \emph{Physical Review E}. {\bf 73}\penalty0 (5),
  \penalty0 056104,  (2006).

\bibitem{jagla2014}
E.~A. Jagla, F.~P. Landes, and A.~Rosso, Viscoelastic {{Effects}} in
  {{Avalanche Dynamics}}: {{A Key}} to {{Earthquake Statistics}}, \emph{Phys.
  Rev. Lett.} {\bf 112}\penalty0 (17),  (2014).
\newblock \doi{10.1103/PhysRevLett.112.174301}.

\bibitem{dearcangelis2016}
L.~{de Arcangelis}, C.~Godano, J.~R. Grasso, and E.~Lippiello, Statistical
  physics approach to earthquake occurrence and forecasting, \emph{Physics
  Reports}. {\bf 628}, \penalty0 1--91,  (2016).
\newblock \doi{10.1016/j.physrep.2016.03.002}.

\bibitem{Sethna07Crossover}
J.~P. Sethna, Statistical mechanics - {C}rackling crossover, \emph{Nature
  Physics (News and Views)}. {\bf 3}, \penalty0 518--519,  (2007).

\bibitem{DurinZapperi06}
G.~Durin and S.~Zapperi.
\newblock The science of hysteresis: Physical modeling, micromagnetics and
  magnetization dynamics, vol. ii, ch. iii (the barkhausen effect),  (2006).

\bibitem{Raindrops}
V.~Berejnov and R.~E. Thorne, Effect of transient pinning on stability of drops
  sitting on an inclined plane, \emph{Phys. Rev. E}. {\bf 75}, \penalty0 066308
  (Jun, 2007).
\newblock \doi{10.1103/PhysRevE.75.066308}.
\newblock URL \url{https://link.aps.org/doi/10.1103/PhysRevE.75.066308}.

\bibitem{Sethna07Wire}
J.~P. Sethna, Crackling wires, \emph{Science (Perspective)}. {\bf 318},
  \penalty0 207--208,  (2007).

\bibitem{DahmenNanopillars}
N.~Friedman, A.~T. Jennings, G.~Tsekenis, J.-Y. Kim, M.~Tao, J.~T. Uhl, J.~R.
  Greer, and K.~A. Dahmen, Statistics of dislocation slip avalanches in
  nanosized single crystals show tuned critical behavior predicted by a simple
  mean field model, \emph{Physical review letters}. {\bf 109}\penalty0 (9),
  \penalty0 095507,  (2012).

\bibitem{SethnaBDGGHKLLNQRRSZ17}
J.~P. {Sethna}, M.~K. {Bierbaum}, K.~A. {Dahmen}, C.~P. {Goodrich}, J.~R.
  {Greer}, L.~X. {Hayden}, J.~P. {Kent-Dobias}, E.~D. {Lee}, D.~B. {Liarte},
  X.~{Ni}, K.~N. {Quinn}, A.~{Raju}, D.~{Zeb Rocklin}, A.~{Shekhawat}, and
  S.~{Zapperi}, Deformation of crystals: Connections with statistical physics,
  \emph{Annual Review of Materials Research}. {\bf 47}, \penalty0 217--246,
  (2017).
\newblock \doi{10.1146/annurev-matsci-070115-032036}.
\newblock URL
  \url{http://www.annualreviews.org/doi/full/10.1146/annurev-matsci-070115-032036}.

\bibitem{NiZLMDSG19}
X.~Ni, H.~Zhang, D.~B. Liarte, L.~W. McFaul, K.~A. Dahmen, J.~P. Sethna, and
  J.~R. Greer, Yield precursor dislocation avalanches in small crystals: The
  irreversibility transition, \emph{Phys. Rev. Lett.} {\bf 123}, \penalty0
  035501 (Jul, 2019).
\newblock \doi{10.1103/PhysRevLett.123.035501}.
\newblock URL \url{https://link.aps.org/doi/10.1103/PhysRevLett.123.035501}.

\bibitem{PapanikolaouDCSUWZ12}
S.~Papanikolaou, D.~M. Dimiduk, W.~Choi, J.~P. Sethna, M.~D. Uchic, C.~F.
  Woodward, and S.~Zapperi, Quasi-periodic events in crystal plasticity and the
  self-organized avalanche oscillator, \emph{Nature}. {\bf 490}\penalty0
  (7421), \penalty0 517--521 (10, 2012).
\newblock URL \url{http://dx.doi.org/10.1038/nature11568}.

\bibitem{Csikor07}
F.~F. Csikor, C.~Motz, D.~Weygand, M.~Zaiser, and S.~Zapperi, Dislocation
  avalanches, strain bursts, and the problem of plastic forming at the
  micrometer scale, \emph{Science}. {\bf 318}\penalty0 (5848), \penalty0
  251--254,  (2007).

\bibitem{ShekhawatZS13}
A.~Shekhawat, S.~Zapperi, and J.~P. Sethna, From damage percolation to crack
  nucleation through finite-size criticality, \emph{Physical Review Letters}.
  {\bf 110}, \penalty0 185505,  (2013).

\bibitem{TewariSDKLL99}
S.~Tewari, D.~Schiemann, D.~J. Durian, C.~M. Knobler, S.~A. Langer, and A.~J.
  Liu, Statistics of shear-induced rearrangements in a two-dimensional model
  foam, \emph{Physical Review E}. {\bf 60}\penalty0 (4), \penalty0 4385,
  (1999).

\bibitem{CieplakR88}
M.~Cieplak and M.~O. Robbins, Dynamical transition in quasistatic fluid
  invasion in porous media, \emph{Physical review letters}. {\bf 60}\penalty0
  (20), \penalty0 2042,  (1988).

\bibitem{ortin2017}
J.~Ort{\'i}n and S.~Santucci.
\newblock Avalanches, {{Non}}-{{Gaussian Fluctuations}} and {{Intermittency}}
  in {{Fluid Imbibition}}.
\newblock In eds. E.~K. Salje, A.~Saxena, and A.~Planes, \emph{Avalanches in
  {{Functional Materials}} and {{Geophysics}}}, pp. 261--292. {Springer
  International Publishing}, {Cham},  (2017).
\newblock ISBN 978-3-319-45610-2 978-3-319-45612-6.
\newblock \doi{10.1007/978-3-319-45612-6_12}.

\bibitem{url:CracklingKids}
M.~Kuntz, P.~Houle, and J.~P. Sethna.
\newblock Crackling noise.
\newblock \url{http://SimScience.org/crackling/},  (1998).

\bibitem{AlavaCoffeeAvalanches}
M.~Alava, M.~Dubé, and M.~Rost, Imbibition in disordered media, \emph{Advances
  in Physics}. {\bf 53}\penalty0 (2), \penalty0 83--175,  (2004).
\newblock \doi{10.1080/00018730410001687363}.
\newblock URL \url{https://doi.org/10.1080/00018730410001687363}.

\bibitem{StarQuakes}
D.~Pines, J.~Shaham, M.~A. Alpar, and P.~W. Anderson, {Pinned Vorticity in
  Rotating Superfluids, with Application to Neutron Stars†)}, \emph{Progress
  of Theoretical Physics Supplement}. {\bf 69}, \penalty0 376--396 (03, 1980).
\newblock ISSN 0375-9687.
\newblock \doi{10.1143/PTP.69.376}.
\newblock URL \url{https://doi.org/10.1143/PTP.69.376}.

\bibitem{WarAvalanches}
E.~D. Lee, B.~C. Daniels, C.~R. Myers, D.~C. Krakauer, and J.~C. Flack, Scaling
  theory of armed-conflict avalanches, \emph{Phys. Rev. E}. {\bf 102},
  \penalty0 042312 (Oct, 2020).
\newblock \doi{10.1103/PhysRevE.102.042312}.
\newblock URL \url{https://link.aps.org/doi/10.1103/PhysRevE.102.042312}.

\bibitem{DahmenNeurons}
N.~Friedman, S.~Ito, B.~A. Brinkman, M.~Shimono, R.~L. DeVille, K.~A. Dahmen,
  J.~M. Beggs, and T.~C. Butler, Universal critical dynamics in high resolution
  neuronal avalanche data, \emph{Physical review letters}. {\bf 108}\penalty0
  (20), \penalty0 208102,  (2012).

\bibitem{beggsP03}
J.~M. Beggs and D.~Plenz, Neuronal avalanches in neocortical circuits,
  \emph{Journal of neuroscience}. {\bf 23}\penalty0 (35), \penalty0
  11167--11177,  (2003).

\bibitem{AlavaPaperTearing}
L.~Salminen, A.~Tolvanen, and M.~J. Alava, Acoustic emission from paper
  fracture, \emph{Physical Review Letters}. {\bf 89}\penalty0 (18), \penalty0
  185503,  (2002).

\bibitem{HouleS96}
P.~A. Houle and J.~P. Sethna, Acoustic emission from crumpling paper,
  \emph{Physical Review E}. {\bf 54}, \penalty0 278--283,  (1996).

\bibitem{KramerL96}
E.~M. Kramer and A.~E. Lobkovsky, Universal power law in the noise from a
  crumpled elastic sheet, \emph{Physical Review E}. {\bf 53}\penalty0 (2),
  \penalty0 1465,  (1996).

\bibitem{MichardB05}
Q.~Michard and J.~Bouchaud, Theory of collective opinion shifts: from smooth
  trends to abrupt swings, \emph{Eur. Phys. J. B}. {\bf 47}, \penalty0
  151–159,  (2005).
\newblock URL \url{https://doi.org/10.1140/epjb/e2005-00307-0}.

\bibitem{bienayme1845}
I.-J. Bienaym{\'e}, De la loi de multiplication et de la dur\'ee des familles,
  \emph{Oc Philomat Paris Extr. S\'er}. {\bf 5}\penalty0 (37-39), \penalty0 4,
  (1845).

\bibitem{watson1875}
H.~W. Watson and F.~Galton, On the {{Probability}} of the {{Extinction}} of
  {{Families}}., \emph{The Journal of the Anthropological Institute of Great
  Britain and Ireland}. {\bf 4}, \penalty0 138,  (1875).
\newblock \doi{10.2307/2841222}.

\bibitem{SethnaDKKRS93}
J.~P. Sethna, K.~Dahmen, S.~Kartha, J.~A. Krumhansl, B.~W. Roberts, and J.~D.
  Shore, Hysteresis and hierarchies - dynamics of disorder-driven 1st-order
  phase-transformations, \emph{Physical Review Letters}. {\bf 70}, \penalty0
  3347--3350,  (1993).

\bibitem{Sethna22}
J.~P. Sethna, Power laws in physics, \emph{Nature Reviews Physics}. {\bf 4},
  \penalty0 501--503 (Jul, 2022).
\newblock \doi{10.1038/s42254-022-00491-x}.
\newblock URL \url{https://rdcu.be/cSbCZ}.

\bibitem{ChenZS15}
Y.-J. Chen, S.~Zapperi, and J.~P. Sethna, Crossover behavior in interface
  depinning, \emph{Phys. Rev. E}. {\bf 92}, \penalty0 022146,  (2015).
\newblock \doi{http://dx.doi.org/10.1103/PhysRevE.92.022146}.
\newblock URL \url{http://link.aps.org/doi/10.1103/PhysRevE.92.022146}.

\bibitem{bak1987self}
P.~Bak, C.~Tang, and K.~Wiesenfeld, Self-organized criticality: An explanation
  of the 1/f noise, \emph{Physical review letters}. {\bf 59}\penalty0 (4),
  \penalty0 381,  (1987).

\bibitem{ZapperiPandemicSOC}
S.~Zapperi, K.~B. Lauritsen, and H.~E. Stanley, Self-organized branching
  processes: Mean-field theory for avalanches, \emph{Phys. Rev. Lett.} {\bf
  75}, \penalty0 4071--4074 (Nov, 1995).
\newblock \doi{10.1103/PhysRevLett.75.4071}.
\newblock URL \url{https://link.aps.org/doi/10.1103/PhysRevLett.75.4071}.

\bibitem{Fisher98}
D.~S. Fisher, Collective transport in random media: from superconductors to
  earthquakes, \emph{Physics Reports}. {\bf 301}\penalty0 (1-3), \penalty0 113
  -- 150,  (1998).
\newblock ISSN 0370-1573.
\newblock \doi{10.1016/S0370-1573(98)00008-8}.
\newblock URL
  \url{http://www.sciencedirect.com/science/article/pii/S0370157398000088}.

\bibitem{kardar1998}
M.~Kardar, Nonequilibrium dynamics of interfaces and lines, \emph{Physics
  Reports}. {\bf 301}\penalty0 (1-3), \penalty0 85--112,  (1998).
\newblock \doi{10.1016/S0370-1573(98)00007-6}.

\bibitem{larkin1970}
A.~Larkin, Model for pinning of vortex lattices, \emph{Sov. Phys. JETP}. {\bf
  31}, \penalty0 784,  (1970).

\bibitem{fisher1985}
D.~S. Fisher, Sliding charge-density waves as a dynamic critical phenomenon,
  \emph{Phys. Rev. B}. {\bf 31}\penalty0 (3), \penalty0 1396--1427,  (1985).
\newblock \doi{10.1103/PhysRevB.31.1396}.

\bibitem{middleton1992a}
A.~A. Middleton, Asymptotic uniqueness of the sliding state for charge-density
  waves, \emph{Phys. Rev. Lett.} {\bf 68}\penalty0 (5), \penalty0 670--673,
  (1992).
\newblock \doi{10.1103/PhysRevLett.68.670}.

\bibitem{giamarchi2006}
T.~Giamarchi, A.~Kolton, and A.~Rosso.
\newblock Dynamics of {{Disordered Elastic Systems}}.
\newblock In eds. M.~C. Miguel and M.~Rubi, \emph{Jamming, {{Yielding}}, and
  {{Irreversible Deformation}} in {{Condensed Matter}}}, vol. 688, pp. 91--108.
  {Springer-Verlag}, {Berlin/Heidelberg},  (2006).
\newblock ISBN 978-3-540-30028-1.
\newblock \doi{10.1007/3-540-33204-9_6}.

\bibitem{joerg2021theory}
K.~J. Wiese, Theory and experiments for disordered elastic manifolds,
  depinning, avalanches, and sandpiles, \emph{Reports on Progress in Physics}.
  {\bf 85}\penalty0 (8), \penalty0 086502,  (2022).

\bibitem{alessandro1990}
B.~Alessandro, C.~Beatrice, G.~Bertotti, and A.~Montorsi, Domain-wall dynamics
  and {{Barkhausen}} effect in metallic ferromagnetic materials. {{I}}.
  {{Theory}}, \emph{Journal of Applied Physics}. {\bf 68}\penalty0 (6),
  \penalty0 2901--2907,  (1990).
\newblock \doi{10.1063/1.346423}.

\bibitem{colaiori2008}
F.~Colaiori, Exactly solvable model of avalanches dynamics for {{Barkhausen}}
  crackling noise, \emph{Advances in Physics}. {\bf 57}\penalty0 (4), \penalty0
  287--359,  (2008).
\newblock \doi{10.1080/00018730802420614}.

\bibitem{ledoussal2009c}
P.~Le~Doussal and K.~J. Wiese, Driven particle in a random landscape:
  {{Disorder}} correlator, avalanche distribution, and extreme value statistics
  of records, \emph{Phys. Rev. E}. {\bf 79}\penalty0 (5), \penalty0 051105,
  (2009).
\newblock \doi{10.1103/PhysRevE.79.051105}.

\bibitem{ledoussal2009b}
P.~Le~Doussal and K.~J. Wiese, Size distributions of shocks and static
  avalanches from the {{Functional Renormalization Group}}, \emph{Phys. Rev.
  E}. {\bf 79}\penalty0 (5), \penalty0 051106,  (2009).
\newblock \doi{10.1103/PhysRevE.79.051106}.

\bibitem{rosso2009}
A.~Rosso, P.~Le~Doussal, and K.~J. Wiese, Avalanche-size distribution at the
  depinning transition: {{A}} numerical test of the theory, \emph{Phys. Rev.
  B}. {\bf 80}\penalty0 (14), \penalty0 144204,  (2009).
\newblock \doi{10.1103/PhysRevB.80.144204}.

\bibitem{le2013avalanche}
P.~Le~Doussal and K.~J. Wiese, Avalanche dynamics of elastic interfaces,
  \emph{Physical Review E}. {\bf 88}\penalty0 (2), \penalty0 022106,  (2013).

\bibitem{delorme2016}
M.~Delorme, P.~Le~Doussal, and K.~Wiese, Distribution of joint local and total
  size and of extension for avalanches in the {{Brownian}} force model,
  \emph{Phys. Rev. E}. {\bf 93},  (2016).
\newblock \doi{10.1103/PhysRevE.93.052142}.

\bibitem{Zoia07}
A.~Zoia, A.~Rosso, and M.~Kardar, Fractional laplacian in bounded domains,
  \emph{Physical Review E}. {\bf 76}\penalty0 (2), \penalty0 021116,  (2007).

\bibitem{joanny1984}
J.~F. Joanny and P.~G. {de Gennes}, A model for contact angle hysteresis,
  \emph{The Journal of Chemical Physics}. {\bf 81}\penalty0 (1), \penalty0
  552--562,  (1984).
\newblock \doi{10.1063/1.447337}.

\bibitem{gao1989}
H.~Gao and J.~Rice, A {{First}}-{{Order Perturbation Analysis}} of {{Crack
  Trapping}} by {{Arrays}} of {{Obstacles}}, \emph{J. Appl. Mech.-Trans. Asme -
  J APPL MECH}. {\bf 56}, \penalty0 828--836,  (1989).
\newblock \doi{10.1115/1.3176178}.

\bibitem{alava2006}
M.~J. Alava, P.~K. V.~V. Nukala, and S.~Zapperi, Statistical models of
  fracture, \emph{Advances in Physics}. {\bf 55}\penalty0 (3-4), \penalty0
  349--476,  (2006).
\newblock \doi{10.1080/00018730300741518}.

\bibitem{bonamy2008}
D.~Bonamy, S.~Santucci, and L.~Ponson, Crackling {{Dynamics}} in {{Material
  Failure}} as the {{Signature}} of a {{Self}}-{{Organized Dynamic Phase
  Transition}}, \emph{Phys. Rev. Lett.} {\bf 101}\penalty0 (4), \penalty0
  045501,  (2008).
\newblock \doi{10.1103/PhysRevLett.101.045501}.

\bibitem{ledoussal2006}
P.~Le~Doussal, K.~J. Wiese, E.~Raphael, and R.~Golestanian, Can {{Nonlinear
  Elasticity Explain Contact}}-{{Line Roughness}} at {{Depinning}}?,
  \emph{Phys. Rev. Lett.} {\bf 96}\penalty0 (1), \penalty0 015702,  (2006).
\newblock \doi{10.1103/PhysRevLett.96.015702}.

\bibitem{bonamy2011}
D.~Bonamy and E.~Bouchaud, Failure of heterogeneous materials: {{A}} dynamic
  phase transition?, \emph{Physics Reports}. {\bf 498}\penalty0 (1), \penalty0
  1--44,  (2011).
\newblock \doi{10.1016/j.physrep.2010.07.006}.

\bibitem{dahmen2009micromechanical}
K.~A. Dahmen, Y.~Ben-Zion, and J.~T. Uhl, Micromechanical model for deformation
  in solids with universal predictions for stress-strain curves and slip
  avalanches, \emph{Physical review letters}. {\bf 102}\penalty0 (17),
  \penalty0 175501,  (2009).

\bibitem{antonaglia2014bulk}
J.~Antonaglia, W.~J. Wright, X.~Gu, R.~R. Byer, T.~C. Hufnagel, M.~LeBlanc,
  J.~T. Uhl, and K.~A. Dahmen, Bulk metallic glasses deform via slip
  avalanches, \emph{Physical review letters}. {\bf 112}\penalty0 (15),
  \penalty0 155501,  (2014).

\bibitem{wright2016experimental}
W.~J. Wright, Y.~Liu, X.~Gu, K.~D. Van~Ness, S.~L. Robare, X.~Liu,
  J.~Antonaglia, M.~LeBlanc, J.~T. Uhl, T.~C. Hufnagel, et~al., Experimental
  evidence for both progressive and simultaneous shear during quasistatic
  compression of a bulk metallic glass, \emph{Journal of Applied Physics}. {\bf
  119}\penalty0 (8), \penalty0 084908,  (2016).

\bibitem{wright2018slip}
W.~J. Wright, A.~A. Long, X.~Gu, X.~Liu, T.~C. Hufnagel, and K.~A. Dahmen, Slip
  statistics for a bulk metallic glass composite reflect its ductility,
  \emph{Journal of Applied Physics}. {\bf 124}\penalty0 (18), \penalty0 185101,
   (2018).

\bibitem{Herschel1926}
W.~H. Herschel and R.~Bulkley, Konsistenzmessungen von
  gummi-benzoll{\"o}sungen, \emph{Kolloid-Zeitschrift}. {\bf 39}\penalty0 (4),
  \penalty0 291--300 (Aug, 1926).
\newblock ISSN 1435-1536.
\newblock \doi{10.1007/BF01432034}.
\newblock URL \url{https://doi.org/10.1007/BF01432034}.

\bibitem{ovarlez2013rheopexy}
G.~Ovarlez, L.~Tocquer, F.~Bertrand, and P.~Coussot, Rheopexy and tunable yield
  stress of carbon black suspensions, \emph{Soft Matter}. {\bf 9}\penalty0
  (23), \penalty0 5540--5549,  (2013).

\bibitem{Argon79}
A.~Argon, Plastic deformation in metallic glasses, \emph{Acta Metallurgica}.
  {\bf 27}\penalty0 (1), \penalty0 47 -- 58,  (1979).
\newblock ISSN 0001-6160.
\newblock \doi{10.1016/0001-6160(79)90055-5}.
\newblock URL
  \url{http://www.sciencedirect.com/science/article/pii/0001616079900555}.

\bibitem{Maloney04}
C.~Maloney and A.~Lemaitre, Subextensive scaling in the athermal, quasistatic
  limit of amorphous matter in plastic shear flow, \emph{Phys. Rev. Lett.} {\bf
  93}, \penalty0 016001 (Jul, 2004).
\newblock \doi{10.1103/PhysRevLett.93.016001}.
\newblock URL \url{http://link.aps.org/doi/10.1103/PhysRevLett.93.016001}.

\bibitem{Salerno12}
K.~M. Salerno, C.~E. Maloney, and M.~O. Robbins, Avalanches in strained
  amorphous solids: Does inertia destroy critical behavior?, \emph{Phys. Rev.
  Lett.} {\bf 109}, \penalty0 105703 (Sep, 2012).
\newblock \doi{10.1103/PhysRevLett.109.105703}.
\newblock URL \url{http://link.aps.org/doi/10.1103/PhysRevLett.109.105703}.

\bibitem{Lin14}
J.~Lin, E.~Lerner, A.~Rosso, and M.~Wyart, Scaling description of the yielding
  transition in soft amorphous solids at zero temperature, \emph{Proceedings of
  the National Academy of Sciences}. {\bf 111}\penalty0 (40), \penalty0
  14382--14387,  (2014).

\bibitem{Baret02}
J.-C. Baret, D.~Vandembroucq, and S.~Roux, Extremal model for amorphous media
  plasticity, \emph{Phys. Rev. Lett.} {\bf 89}, \penalty0 195506 (Oct, 2002).
\newblock \doi{10.1103/PhysRevLett.89.195506}.
\newblock URL \url{http://link.aps.org/doi/10.1103/PhysRevLett.89.195506}.

\bibitem{Picard2004}
G.~Picard, A.~Ajdari, F.~Lequeux, and L.~Bocquet, Elastic consequences of a
  single plastic event: a step towards the microscopic modelling of the flow of
  yield stress fluids, \emph{Eur. Phys. Jour. E}. {\bf 15}, \penalty0 371,
  (2004).

\bibitem{budrikis2017universal}
Z.~Budrikis, D.~F. Castellanos, S.~Sandfeld, M.~Zaiser, and S.~Zapperi,
  Universal features of amorphous plasticity, \emph{Nature communications}.
  {\bf 8}\penalty0 (1), \penalty0 1--10,  (2017).

\bibitem{CarlsonL89}
J.~Carlson and J.~Langer, Mechanical model of an earthquake fault,
  \emph{Physical Review A}. {\bf 40}\penalty0 (11), \penalty0 6470,  (1989).

\bibitem{nicolas2018deformation}
A.~Nicolas, E.~E. Ferrero, K.~Martens, and J.-L. Barrat, Deformation and flow
  of amorphous solids: Insights from elastoplastic models, \emph{Reviews of
  Modern Physics}. {\bf 90}\penalty0 (4), \penalty0 045006,  (2018).

\bibitem{cao2018a}
X.~Cao, A.~Nicolas, D.~Trimcev, and A.~Rosso, Soft modes and strain
  redistribution in continuous models of amorphous plasticity: The {{Eshelby}}
  paradigm, and beyond?, \emph{Soft Matter}. {\bf 14}\penalty0 (18), \penalty0
  3640--3651,  (2018).
\newblock \doi{10.1039/C7SM02510F}.

\bibitem{PedestrianJamming}
M.~Muramatsu, T.~Irie, and T.~Nagatani, Jamming transition in pedestrian
  counter flow, \emph{Physica A: Statistical Mechanics and its Applications}.
  {\bf 267}\penalty0 (3-4), \penalty0 487--498,  (1999).

\bibitem{WyartNeuralNetworkJamming}
M.~Geiger, L.~Petrini, and M.~Wyart, Landscape and training regimes in deep
  learning, \emph{Physics Reports}. {\bf 924}, \penalty0 1--18,  (2021).

\bibitem{Andreotti13}
B.~Andreotti, Y.~Forterre, and O.~Pouliquen, \emph{Granular media: between
  fluid and solid}. (Cambridge University Press, 2013).

\bibitem{nowak2005maximum}
S.~Nowak, A.~Samadani, and A.~Kudrolli, Maximum angle of stability of a wet
  granular pile, \emph{Nature Physics}. {\bf 1}\penalty0 (1), \penalty0 50--52,
   (2005).

\bibitem{perrin2019interparticle}
H.~Perrin, C.~Clavaud, M.~Wyart, B.~Metzger, and Y.~Forterre, Interparticle
  friction leads to nonmonotonic flow curves and hysteresis in viscous
  suspensions, \emph{Physical Review X}. {\bf 9}\penalty0 (3), \penalty0
  031027,  (2019).

\bibitem{Dijksman11}
J.~A. Dijksman, G.~H. Wortel, L.~T. van Dellen, O.~Dauchot, and M.~van Hecke,
  Jamming, yielding, and rheology of weakly vibrated granular media,
  \emph{Physical review letters}. {\bf 107}\penalty0 (10), \penalty0 108303,
  (2011).

\bibitem{mowlavi2021interplay}
S.~Mowlavi and K.~Kamrin, Interplay between hysteresis and nonlocality during
  onset and arrest of flow in granular materials, \emph{Soft Matter}. {\bf
  17}\penalty0 (31), \penalty0 7359--7375,  (2021).

\bibitem{DeGiuli17a}
E.~DeGiuli and M.~Wyart, Friction law and hysteresis in granular materials,
  \emph{Proceedings of the National Academy of Sciences}. {\bf 114}\penalty0
  (35), \penalty0 9284--9289,  (2017).
\newblock URL \url{http://www.pnas.org/content/114/35/9284.abstract}.

\bibitem{Pouliquen04}
O.~Pouliquen, Velocity correlations in dense granular flows, \emph{Physical
  review letters}. {\bf 93}\penalty0 (24), \penalty0 248001,  (2004).

\bibitem{Olsson10}
P.~Olsson, Asymmetric velocity correlations in shearing media, \emph{Phys. Rev.
  E}. {\bf 82}\penalty0 (3), \penalty0 031303 (Sep, 2010).
\newblock \doi{10.1103/PhysRevE.82.031303}.

\bibitem{During14}
G.~D{\"u}ring, E.~Lerner, and M.~Wyart, Length scales and self-organization in
  dense suspension flows, \emph{Physical Review E}. {\bf 89}\penalty0 (2),
  \penalty0 022305,  (2014).

\bibitem{Olsson11}
P.~Olsson and S.~Teitel, Critical scaling of shearing rheology at the jamming
  transition of soft-core frictionless disks, \emph{Physical Review E}. {\bf
  83}\penalty0 (3), \penalty0 030302,  (2011).

\bibitem{degiuli14d}
E.~DeGiuli, G.~D{\"u}ring, E.~Lerner, and M.~Wyart, Unified theory of inertial
  granular flows and non-{B}rownian suspensions, \emph{arXiv preprint
  arXiv:1410.3535}.  (2014).

\bibitem{perrin2021nonlocal}
H.~Perrin, M.~Wyart, B.~Metzger, and Y.~Forterre, Nonlocal effects reflect the
  jamming criticality in frictionless granular flows down inclines,
  \emph{Physical Review Letters}. {\bf 126},  (2021).

\bibitem{Liu10}
A.~J. Liu, S.~R. Nagel, W.~van Saarloos, and M.~Wyart, \emph{The jamming
  scenario: an introduction and outlook}, In eds. L.Berthier, G.~Biroli,
  J.~Bouchaud, L.~Cipeletti, and W.~van Saarloos, \emph{Dynamical
  heterogeneities in glasses, colloids, and granular media}, p. 298.
\newblock Oxford University Press, Oxford,  (2010).

\bibitem{Hecke10}
M.~van Hecke, Jamming of soft particles: geometry, mechanics, scaling and
  isostaticity, \emph{Journal of Physics: Condensed Matter}. {\bf 22}\penalty0
  (3), \penalty0 033101--033124,  (2010).

\bibitem{OHern03}
C.~S. O'Hern, L.~E. Silbert, A.~J. Liu, and S.~R. Nagel, Jamming at zero
  temperature and zero applied stress: The epitome of disorder, \emph{Phys.
  Rev. E}. {\bf 68}\penalty0 (1), \penalty0 011306--011324 (Jul, 2003).
\newblock \doi{10.1103/PhysRevE.68.011306}.

\bibitem{Peyneau08}
P.-E. Peyneau and J.-N. Roux, Frictionless bead packs have macroscopic
  friction, but no dilatancy, \emph{Physical review E}. {\bf 78}\penalty0 (1),
  \penalty0 011307,  (2008).

\bibitem{Lespiat11}
R.~Lespiat, S.~Cohen-Addad, and R.~H\"ohler, Jamming and flow of
  random-close-packed spherical bubbles: An analogy with granular materials,
  \emph{Phys. Rev. Lett.} {\bf 106}, \penalty0 148302 (Apr, 2011).
\newblock \doi{10.1103/PhysRevLett.106.148302}.

\bibitem{Combe00}
G.~Combe and J.-N. Roux, Strain versus stress in a model granular material: a
  devil's staircase, \emph{Physical Review Letters}. {\bf 85}\penalty0 (17),
  \penalty0 3628,  (2000).

\bibitem{Lerner13a}
E.~Lerner, G.~During, and M.~Wyart, Low-energy non-linear excitations in sphere
  packings, \emph{Soft Matter}. {\bf 9}, \penalty0 8252--8263,  (2013).
\newblock \doi{10.1039/C3SM50515D}.

\bibitem{Efros75}
A.~L. Efros and B.~I. Shklovskii, Coulomb gap and low temperature conductivity
  of disordered systems, \emph{Journal of Physics C: Solid State Physics}. {\bf
  8}\penalty0 (4), \penalty0 L49,  (1975).
\newblock URL \url{http://stacks.iop.org/0022-3719/8/i=4/a=003}.

\bibitem{Thouless77}
D.~Thouless, P.~Anderson, and R.~Palmer, Solution of solvable model of a spin
  glass, \emph{Philo. Mag.} {\bf 35}, \penalty0 593--601 (Jul, 1977).
\newblock \doi{10.1080/14786437708235992}.
\newblock URL \url{http://link.aps.org/doi/10.1103/PhysRevLett.59.381}.

\bibitem{Karmakar10a}
S.~Karmakar, E.~Lerner, and I.~Procaccia, Statistical physics of the yielding
  transition in amorphous solids, \emph{Phys. Rev. E}. {\bf 82}, \penalty0
  055103 (Nov, 2010).
\newblock \doi{10.1103/PhysRevE.82.055103}.
\newblock URL \url{http://link.aps.org/doi/10.1103/PhysRevE.82.055103}.

\bibitem{Lin14a}
J.~Lin, A.~Saade, E.~Lerner, A.~Rosso, and M.~Wyart, On the density of shear
  transformations in amorphous solids, \emph{EPL (Europhysics Letters)}. {\bf
  105}\penalty0 (2), \penalty0 26003,  (2014).

\bibitem{Lin15a}
J.~Lin, T.~Gueudr{\'e}, A.~Rosso, and M.~Wyart, Criticality in the approach to
  failure in amorphous solids, \emph{Phys. Rev. Lett.} {\bf 115}, \penalty0
  168001,  (2015).

\bibitem{Muller14}
M.~M{\"u}ller and M.~Wyart, Marginal stability in structural, spin, and
  electron glasses, \emph{Annual Review of Condensed Matter Physics}. {\bf
  6}\penalty0 (1), \penalty0 177--200,  (2015).
\newblock \doi{10.1146/annurev-conmatphys-031214-014614}.

\bibitem{Maloney06a}
C.~E. Maloney and A.~Lema\^itre, Amorphous systems in athermal, quasistatic
  shear, \emph{Phys. Rev. E}. {\bf 74}\penalty0 (1), \penalty0 016118 (Jul,
  2006).

\bibitem{kane2014topological}
C.~Kane and T.~Lubensky, Topological boundary modes in isostatic lattices,
  \emph{Nature Physics}. {\bf 10}\penalty0 (1), \penalty0 39--45,  (2014).

\bibitem{Wyart12}
M.~Wyart, Marginal stability constrains force and pair distributions at random
  close packing, \emph{Phys. Rev. Lett.} {\bf 109}, \penalty0 125502 (Sep,
  2012).
\newblock \doi{10.1103/PhysRevLett.109.125502}.

\bibitem{Charbonneau14}
P.~Charbonneau, J.~Kurchan, G.~Parisi, P.~Urbani, and F.~Zamponi, Fractal free
  energy landscapes in structural glasses, \emph{Nature Communications}. {\bf
  5}\penalty0 (3725),  (2014).

\bibitem{Charbonneau15}
P.~Charbonneau, E.~I. Corwin, G.~Parisi, and F.~Zamponi, Jamming criticality
  revealed by removing localized buckling excitations, \emph{Physical Review
  Letters}. {\bf 114}\penalty0 (12), \penalty0 125504,  (2015).

\bibitem{GoodrichLiuUCD2}
C.~P. Goodrich, A.~J. Liu, and S.~R. Nagel, Finite-size scaling at the jamming
  transition, \emph{Phys. Rev. Lett.} {\bf 109}, \penalty0 095704 (Aug, 2012).
\newblock \doi{10.1103/PhysRevLett.109.095704}.
\newblock URL \url{https://link.aps.org/doi/10.1103/PhysRevLett.109.095704}.

\bibitem{SartorRCnn}
J.~D. Sartor, S.~A. Ridout, and E.~I. Corwin, Mean-field predictions of scaling
  prefactors match low-dimensional jammed packings, \emph{Physical Review
  Letters}. {\bf 126}\penalty0 (4), \penalty0 048001,  (2021).

\bibitem{behrens2022dis}
F.~Behrens, G.~Arpino, Y.~Kivva, and L.~Zdeborov{\'a}, (dis) assortative
  partitions on random regular graphs, \emph{arXiv preprint arXiv:2202.10379}.
  (2022).

\bibitem{Lemaitre07a}
A.~Lema{\^\i}tre and C.~Caroli, Plastic response of a two-dimensional amorphous
  solid to quasistatic shear: Transverse particle diffusion and phenomenology
  of dissipative events, \emph{Physical Review E}. {\bf 76}\penalty0 (3),
  \penalty0 036104,  (2007).

\bibitem{nattermann1990}
T.~Nattermann, Y.~Shapir, and I.~Vilfan, Interface pinning and dynamics in
  random systems, \emph{Phys. Rev. B}. {\bf 42}\penalty0 (13), \penalty0
  8577--8586,  (1990).
\newblock \doi{10.1103/PhysRevB.42.8577}.

\bibitem{agoritsas2016driven}
E.~Agoritsas, R.~Garc{\'\i}a-Garc{\'\i}a, V.~Lecomte, L.~Truskinovsky, and
  D.~Vandembroucq, Driven interfaces: from flow to creep through model
  reduction, \emph{Journal of Statistical Physics}. {\bf 164}\penalty0 (6),
  \penalty0 1394--1428,  (2016).

\bibitem{tanguy1998}
A.~Tanguy, M.~Gounelle, and S.~Roux, From individual to collective pinning:
  {{Effect}} of long-range elastic interactions, \emph{Phys. Rev. E}. {\bf
  58}\penalty0 (2), \penalty0 1577--1590,  (1998).
\newblock \doi{10.1103/PhysRevE.58.1577}.

\bibitem{chauve2000}
P.~Chauve, T.~Giamarchi, and P.~Le~Doussal, Creep and depinning in disordered
  media, \emph{Phys. Rev. B}. {\bf 62}\penalty0 (10), \penalty0 6241--6267,
  (2000).
\newblock \doi{10.1103/PhysRevB.62.6241}.

\bibitem{cao2018}
X.~Cao, S.~Bouzat, A.~B. Kolton, and A.~Rosso, Localization of soft modes at
  the depinning transition, \emph{Phys. Rev. E}. {\bf 97}\penalty0 (2),
  (2018).
\newblock \doi{10.1103/PhysRevE.97.022118}.

\bibitem{narayan1992}
O.~Narayan and D.~S. Fisher, Critical behavior of sliding charge-density waves
  in 4-{$\epsilon$} dimensions, \emph{Phys. Rev. B}. {\bf 46}\penalty0 (18),
  \penalty0 11520--11549,  (1992).
\newblock \doi{10.1103/PhysRevB.46.11520}.

\bibitem{nattermann1992}
T.~Nattermann and L.-H. Tang, Kinetic surface roughening. {{I}}. {{The
  Kardar}}-{{Parisi}}-{{Zhang}} equation in the weak-coupling regime,
  \emph{Phys. Rev. A}. {\bf 45}\penalty0 (10), \penalty0 7156--7161,  (1992).
\newblock \doi{10.1103/PhysRevA.45.7156}.

\bibitem{narayan1993}
O.~Narayan and D.~S. Fisher, Threshold critical dynamics of driven interfaces
  in random media, \emph{Phys. Rev. B}. {\bf 48}\penalty0 (10), \penalty0
  7030--7042,  (1993).
\newblock \doi{10.1103/PhysRevB.48.7030}.

\bibitem{ertas1994}
D.~Erta{\c s} and M.~Kardar, Critical dynamics of contact line depinning,
  \emph{Phys. Rev. E}. {\bf 49}\penalty0 (4), \penalty0 R2532--R2535,  (1994).
\newblock \doi{10.1103/PhysRevE.49.R2532}.

\bibitem{chauve2001}
P.~Chauve, P.~Le~Doussal, and K.~J{\"o}rg~Wiese, Renormalization of {{Pinned
  Elastic Systems}}: {{How Does It Work Beyond One Loop}}?, \emph{Phys. Rev.
  Lett.} {\bf 86}\penalty0 (9), \penalty0 1785--1788,  (2001).
\newblock \doi{10.1103/PhysRevLett.86.1785}.

\bibitem{ledoussal2002}
P.~Le~Doussal, K.~J. Wiese, and P.~Chauve, 2-loop {{Functional Renormalization
  Group Theory}} of the {{Depinning Transition}}, \emph{Phys. Rev. B}. {\bf
  66}\penalty0 (17),  (2002).
\newblock \doi{10.1103/PhysRevB.66.174201}.

\bibitem{ledoussal2004}
P.~Le~Doussal, K.~J. Wiese, and P.~Chauve, Functional {{Renormalization Group}}
  and the {{Field Theory}} of {{Disordered Elastic Systems}}, \emph{Phys. Rev.
  E}. {\bf 69}\penalty0 (2),  (2004).
\newblock \doi{10.1103/PhysRevE.69.026112}.

\bibitem{wiese2018field}
K.~J. Wiese, C.~Husemann, and P.~Le~Doussal, Field theory of disordered elastic
  interfaces at 3-loop order: The $\beta$-function, \emph{Nuclear Physics B}.
  {\bf 932}, \penalty0 540--588,  (2018).

\bibitem{husemann2018field}
C.~Husemann and K.~J. Wiese, Field theory of disordered elastic interfaces at
  3-loop order: Critical exponents and scaling functions, \emph{Nuclear Physics
  B}. {\bf 932}, \penalty0 589--618,  (2018).

\bibitem{wiese2021force}
C.~ter Burg, G.~Durin, and K.~J. Wiese, Force-force correlations in disordered
  magnets, \emph{arXiv preprint arXiv:2109.01197}.  (2021).

\bibitem{leschhorn1997}
H.~Leschhorn, T.~Nattermann, S.~Stepanow, and L.-H. Tang, {Driven interface
  depinning in a disordered medium}, \emph{Ann. Phys.} {\bf 509}\penalty0 (1),
  \penalty0 1--34,  (1997).
\newblock \doi{10.1002/andp.19975090102}.

\bibitem{rosso2002}
A.~Rosso and W.~Krauth, Roughness at the depinning threshold for a long-range
  elastic string, \emph{Phys. Rev. E}. {\bf 65}\penalty0 (2), \penalty0 025101,
   (2002).
\newblock \doi{10.1103/PhysRevE.65.025101}.

\bibitem{rosso2003}
A.~Rosso, A.~K. Hartmann, and W.~Krauth, Depinning of elastic manifolds,
  \emph{Phys. Rev. E}. {\bf 67}\penalty0 (2), \penalty0 021602,  (2003).
\newblock \doi{10.1103/PhysRevE.67.021602}.

\bibitem{duemmer2005}
O.~Duemmer and W.~Krauth, Critical exponents of the driven elastic string in a
  disordered medium, \emph{Phys. Rev. E}. {\bf 71}\penalty0 (6),  (2005).
\newblock \doi{10.1103/PhysRevE.71.061601}.

\bibitem{duemmer2007}
O.~Duemmer and W.~Krauth, Depinning exponents of the driven long-range elastic
  string, \emph{J. Stat. Mech. Theory Exp.} {\bf 2007}\penalty0 (01), \penalty0
  P01019--P01019,  (2007).
\newblock \doi{10.1088/1742-5468/2007/01/P01019}.

\bibitem{ferrero2013}
E.~E. Ferrero, S.~Bustingorry, A.~B. Kolton, and A.~Rosso, Numerical
  {{Approaches}} on {{Driven Elastic Interfaces}} in {{Random Media}},
  \emph{Comptes Rendus Phys.} {\bf 14}\penalty0 (8), \penalty0 641--650,
  (2013).
\newblock \doi{10.1016/j.crhy.2013.08.002}.

\bibitem{rosso2007}
A.~Rosso, P.~L. Doussal, and K.~J. Wiese, Numerical {{Calculation}} of the
  {{Functional}} renormalization group fixed-point functions at the depinning
  transition, \emph{Phys. Rev. B}. {\bf 75}\penalty0 (22), \penalty0 220201,
  (2007).
\newblock \doi{10.1103/PhysRevB.75.220201}.

\bibitem{lemerle1998}
S.~Lemerle, J.~Ferr{\'e}, C.~Chappert, V.~Mathet, T.~Giamarchi, and
  P.~Le~Doussal, Domain {{Wall Creep}} in an {{Ising Ultrathin Magnetic Film}},
  \emph{Phys. Rev. Lett.} {\bf 80}\penalty0 (4), \penalty0 849--852,  (1998).
\newblock \doi{10.1103/PhysRevLett.80.849}.

\bibitem{caballero2018}
N.~B. Caballero, E.~E. Ferrero, A.~B. Kolton, J.~Curiale, V.~Jeudy, and
  S.~Bustingorry, Magnetic domain wall creep and depinning: {{A}} scalar field
  model approach, \emph{Phys. Rev. E}. {\bf 97}\penalty0 (6), \penalty0 062122,
   (2018).
\newblock \doi{10.1103/PhysRevE.97.062122}.

\bibitem{ferrero2020}
E.~E. Ferrero, L.~Foini, T.~Giamarchi, A.~B. Kolton, and A.~Rosso, Creep motion
  of elastic interfaces driven in a disordered landscape, \emph{ArXiv200111464
  Cond-Mat}.  (2020).

\bibitem{laurson2009}
L.~Laurson, X.~Illa, and M.~J. Alava, The effect of thresholding on temporal
  avalanche statistics, \emph{J. Stat. Mech.} {\bf 2009}\penalty0 (01),
  \penalty0 P01019,  (2009).
\newblock \doi{10.1088/1742-5468/2009/01/P01019}.

\bibitem{bares2019}
J.~Bar{\'e}s, D.~Bonamy, and A.~Rosso, Seismiclike organization of avalanches
  in a driven long-range elastic string as a paradigm of brittle cracks,
  \emph{Phys. Rev. E}. {\bf 100}\penalty0 (2), \penalty0 023001,  (2019).
\newblock \doi{10.1103/PhysRevE.100.023001}.

\bibitem{bolech2004}
C.~J. Bolech and A.~Rosso, Universal {{Statistics}} of the {{Critical Depinning
  Force}} of {{Elastic Systems}} in {{Random Media}}, \emph{Phys. Rev. Lett.}
  {\bf 93}\penalty0 (12), \penalty0 125701,  (2004).
\newblock \doi{10.1103/PhysRevLett.93.125701}.

\bibitem{demery2014}
V.~D{\'e}mery, A.~Rosso, and L.~Ponson, From microstructural features to
  effective toughness in disordered brittle solids, \emph{EPL Europhys. Lett.}
  {\bf 105}\penalty0 (3), \penalty0 34003,  (2014).
\newblock \doi{10.1209/0295-5075/105/34003}.

\bibitem{fedorenko2006}
A.~A. Fedorenko, P.~Le~Doussal, and K.~J. Wiese, Universal distribution of
  threshold forces at the depinning transition, \emph{Phys. Rev. E}. {\bf
  74}\penalty0 (4), \penalty0 041110,  (2006).
\newblock \doi{10.1103/PhysRevE.74.041110}.

\bibitem{PapanikolaouBSDZS11}
S.~Papanikolaou, F.~Bohn, R.~L. Sommer, G.~Durin, S.~Zapperi, and J.~P. Sethna,
  Universality beyond power laws and the average avalanche shape, \emph{Nature
  Physics}. {\bf 7}, \penalty0 316--320,  (2011).
\newblock \doi{10.1038/NPHYS1884}.

\bibitem{KuntzS00}
M.~C. Kuntz and J.~P. Sethna, Noise in disordered systems: The power spectrum
  and dynamic exponents in avalanche models, \emph{Physical Review B}. {\bf
  62}, \penalty0 11699--11708,  (2000).

\bibitem{MehtaMDS02}
A.~P. Mehta, A.~C. Mills, K.~A. Dahmen, and J.~P. Sethna, Universal pulse shape
  scaling function and exponents: Critical test for avalanche models applied to
  {B}arkhausen noise, \emph{Physical Review E}. {\bf 65}, \penalty0 046139,
  (2002).

\bibitem{ZapperiCCD05}
S.~Zapperi, C.~Castellano, F.~Colaiori, and G.~Durin, Signature of effective
  mass in crackling-noise asymmetry, \emph{Nature Physics}. {\bf 1}\penalty0
  (1), \penalty0 46--49,  (2005).

\bibitem{LeDoussalWieseAverageTemporalShape}
P.~Le~Doussal and K.~J. Wiese, Distribution of velocities in an avalanche,
  \emph{EPL (Europhysics Letters)}. {\bf 97}\penalty0 (4), \penalty0 46004,
  (2012).

\bibitem{Dobrinevski15avalanche}
A.~Dobrinevski, P.~Le~Doussal, and K.~J. Wiese, Avalanche shape and exponents
  beyond mean-field theory, \emph{EPL (Europhysics Letters)}. {\bf
  108}\penalty0 (6), \penalty0 66002,  (2015).

\bibitem{durin2016quantitative}
G.~Durin, F.~Bohn, M.~A. Corr{\^e}a, R.~L. Sommer, P.~Le~Doussal, and K.~Wiese,
  Quantitative scaling of magnetic avalanches, \emph{Physical review letters}.
  {\bf 117}\penalty0 (8), \penalty0 087201,  (2016).

\bibitem{dobrinevski2013statistics}
A.~Dobrinevski, P.~Le~Doussal, and K.~J. Wiese, Statistics of avalanches with
  relaxation and barkhausen noise: A solvable model, \emph{Physical Review E}.
  {\bf 88}\penalty0 (3), \penalty0 032106,  (2013).

\bibitem{lepriol2020}
C.~Le~Priol, J.~Chopin, P.~Le~Doussal, L.~Ponson, and A.~Rosso, Universal
  {{Scaling}} of the {{Velocity Field}} in {{Crack Front Propagation}},
  \emph{Phys. Rev. Lett.} {\bf 124}\penalty0 (6), \penalty0 065501,  (2020).
\newblock \doi{10.1103/PhysRevLett.124.065501}.

\bibitem{maloy2006}
K.~J. M{\aa}l{\o}y, S.~Santucci, J.~Schmittbuhl, and R.~Toussaint, Local
  {{Waiting Time Fluctuations}} along a {{Randomly Pinned Crack Front}},
  \emph{Phys. Rev. Lett.} {\bf 96}\penalty0 (4), \penalty0 045501,  (2006).
\newblock \doi{10.1103/PhysRevLett.96.045501}.

\bibitem{laurson2010}
L.~Laurson, S.~Santucci, and S.~Zapperi, Avalanches and clusters in planar
  crack front propagation, \emph{Phys. Rev. E}. {\bf 81}\penalty0 (4),  (2010).
\newblock \doi{10.1103/PhysRevE.81.046116}.

\bibitem{cao2022clusters}
X.~Cao, P.~L. Doussal, and A.~Rosso, Clusters in an epidemic model with
  long-range dispersal, \emph{arXiv preprint arXiv:2203.14663}.  (2022).

\bibitem{lepriol2021}
C.~Le~Priol, P.~Le~Doussal, and A.~Rosso, Spatial clustering of depinning
  avalanches in presence of long-range interactions, \emph{Physical Review
  Letters}. {\bf 126}\penalty0 (2), \penalty0 025702,  (2021).

\bibitem{Hebraud98}
P.~H\'ebraud and F.~Lequeux, Mode-coupling theory for the pasty rheology of
  soft glassy materials, \emph{Phys. Rev. Lett.} {\bf 81}, \penalty0 2934--2937
  (Oct, 1998).
\newblock \doi{10.1103/PhysRevLett.81.2934}.
\newblock URL \url{http://link.aps.org/doi/10.1103/PhysRevLett.81.2934}.

\bibitem{jagla2015avalanche}
E.~A. Jagla, Avalanche-size distributions in mean-field plastic yielding
  models, \emph{Physical Review E}. {\bf 92}\penalty0 (4), \penalty0 042135,
  (2015).

\bibitem{Lemaitre07}
A.~Lema{\^\i}tre and C.~Caroli, Plastic response of a 2d amorphous solid to
  quasi-static shear: Ii-dynamical noise and avalanches in a mean field model,
  \emph{arXiv preprint arXiv:0705.3122}.  (2007).

\bibitem{Lin16}
J.~Lin and M.~Wyart, Mean-field description of plastic flow in amorphous
  solids, \emph{Physical Review X}. {\bf 6}\penalty0 (1), \penalty0 011005,
  (2016).

\bibitem{ji2019theory}
W.~Ji, M.~Popovi{\'c}, T.~W. de~Geus, E.~Lerner, and M.~Wyart, Theory for the
  density of interacting quasilocalized modes in amorphous solids,
  \emph{Physical Review E}. {\bf 99}\penalty0 (2), \penalty0 023003,  (2019).

\bibitem{ozawa2018random}
M.~Ozawa, L.~Berthier, G.~Biroli, A.~Rosso, and G.~Tarjus, Random critical
  point separates brittle and ductile yielding transitions in amorphous
  materials, \emph{Proceedings of the National Academy of Sciences}. {\bf
  115}\penalty0 (26), \penalty0 6656--6661,  (2018).

\bibitem{lin2018microscopic}
J.~Lin and M.~Wyart, Microscopic processes controlling the herschel-bulkley
  exponent, \emph{Physical review E}. {\bf 97}\penalty0 (1), \penalty0 012603,
  (2018).

\bibitem{parley2022mean}
J.~T. Parley, R.~Mandal, and P.~Sollich, Mean-field description of aging linear
  response in athermal amorphous solids, \emph{Physical Review Materials}. {\bf
  6}\penalty0 (6), \penalty0 065601,  (2022).

\bibitem{Salerno13}
K.~M. Salerno and M.~O. Robbins, Effect of inertia on sheared disordered
  solids: Critical scaling of avalanches in two and three dimensions,
  \emph{Physical Review E}. {\bf 88}\penalty0 (6), \penalty0 062206,  (2013).

\bibitem{popovic2021scaling}
M.~Popovi{\'c}, T.~W. de~Geus, W.~Ji, A.~Rosso, and M.~Wyart, Scaling
  description of creep flow in amorphous solids, \emph{arXiv preprint
  arXiv:2111.04061}.  (2021).

\bibitem{balmforth2014yielding}
N.~J. Balmforth, I.~A. Frigaard, and G.~Ovarlez, Yielding to stress: recent
  developments in viscoplastic fluid mechanics, \emph{Annu. Rev. Fluid Mech}.
  {\bf 46}\penalty0 (1), \penalty0 121--146,  (2014).

\bibitem{goyon2010does}
J.~Goyon, A.~Colin, and L.~Bocquet, How does a soft glassy material flow:
  finite size effects, non local rheology, and flow cooperativity, \emph{Soft
  Matter}. {\bf 6}\penalty0 (12), \penalty0 2668--2678,  (2010).

\bibitem{gueudre2017scaling}
T.~Gueudr{\'e}, J.~Lin, A.~Rosso, and M.~Wyart, Scaling description of
  non-local rheology, \emph{Soft Matter}. {\bf 13}\penalty0 (20), \penalty0
  3794--3801,  (2017).

\bibitem{popovic2021thermally}
M.~Popovi{\'c}, T.~W. de~Geus, W.~Ji, and M.~Wyart, Thermally activated flow in
  models of amorphous solids, \emph{Physical Review E}. {\bf 104}\penalty0 (2),
  \penalty0 025010,  (2021).

\bibitem{ferrero2021yielding}
E.~E. Ferrero, A.~B. Kolton, and E.~A. Jagla, Yielding of amorphous solids at
  finite temperatures, \emph{Physical Review Materials}. {\bf 5}\penalty0 (11),
  \penalty0 115602,  (2021).

\bibitem{Le-Doussal12}
P.~Le~Doussal, M.~M\"uller, and K.~J. Wiese, Equilibrium avalanches in spin
  glasses, \emph{Phys. Rev. B}. {\bf 85}, \penalty0 214402 (Jun, 2012).
\newblock \doi{10.1103/PhysRevB.85.214402}.
\newblock URL \url{http://link.aps.org/doi/10.1103/PhysRevB.85.214402}.

\bibitem{franz2017mean}
S.~Franz and S.~Spigler, Mean-field avalanches in jammed spheres,
  \emph{Physical Review E}. {\bf 95}\penalty0 (2), \penalty0 022139,  (2017).

\bibitem{TarjusNonPerturbativeRG5.1}
I.~Balog, G.~Tarjus, and M.~Tissier, Dimensional reduction breakdown and
  correction to scaling in the random-field {I}sing model, \emph{Phys. Rev. E}.
  {\bf 102}, \penalty0 062154 (Dec, 2020).
\newblock \doi{10.1103/PhysRevE.102.062154}.
\newblock URL \url{https://link.aps.org/doi/10.1103/PhysRevE.102.062154}.

\bibitem{HaydenRS19}
L.~X. Hayden, A.~Raju, and J.~P. Sethna, Unusual scaling for two-dimensional
  avalanches: Curing the faceting and scaling in the lower critical dimension,
  \emph{Phys. Rev. Research}. {\bf 1}, \penalty0 033060 (Oct, 2019).
\newblock \doi{10.1103/PhysRevResearch.1.033060}.
\newblock URL \url{https://link.aps.org/doi/10.1103/PhysRevResearch.1.033060}.

\bibitem{shimada2018spatial}
M.~Shimada, H.~Mizuno, M.~Wyart, and A.~Ikeda, Spatial structure of
  quasilocalized vibrations in nearly jammed amorphous solids, \emph{Physical
  Review E}. {\bf 98}\penalty0 (6), \penalty0 060901,  (2018).

\bibitem{Olsson07}
P.~{Olsson} and S.~{Teitel}, {Critical Scaling of Shear Viscosity at the
  Jamming Transition}, \emph{Phys.\ Rev.\ Lett.} {\bf 99}, \penalty0 178001,
  (2007).

\bibitem{Heussinger09}
C.~Heussinger and J.-L. Barrat, Jamming transition as probed by quasistatic
  shear flow, \emph{Phys. Rev. Lett.} {\bf 102}, \penalty0 218303,  (2009).

\bibitem{Lerner14}
E.~Lerner, E.~DeGiuli, G.~D{\"u}ring, and M.~Wyart, Breakdown of continuum
  elasticity in amorphous solids, \emph{Soft Matter}. {\bf 10}, \penalty0
  5085--5092,  (2014).

\bibitem{olsson2019dimensionality}
P.~Olsson, Dimensionality and viscosity exponent in shear-driven jamming,
  \emph{Physical Review Letters}. {\bf 122}\penalty0 (10), \penalty0 108003,
  (2019).

\bibitem{Trulsson12}
M.~Trulsson, B.~Andreotti, and P.~Claudin, Transition from the viscous to
  inertial regime in dense suspensions, \emph{Physical review letters}. {\bf
  109}\penalty0 (11), \penalty0 118305,  (2012).

\bibitem{DeGiuli15a}
E.~DeGiuli, G.~D\"uring, E.~Lerner, and M.~Wyart, Unified theory of inertial
  granular flows and non-brownian suspensions, \emph{Physical Review E}. {\bf
  91}\penalty0 (6), \penalty0 062206 (06, 2015).

\bibitem{Mari14}
R.~Mari, R.~Seto, J.~F. Morris, and M.~M. Denn, Shear thickening, frictionless
  and frictional rheologies in non-brownian suspensions, \emph{Journal of
  Rheology (1978-present)}. {\bf 58}\penalty0 (6), \penalty0 1693--1724,
  (2014).

\bibitem{Wyart14}
M.~Wyart and M.~Cates, Discontinuous shear thickening without inertia in dense
  non-brownian suspensions, \emph{Physical review letters}. {\bf 112}\penalty0
  (9), \penalty0 098302,  (2014).

\bibitem{clavaud2017revealing}
C.~Clavaud, A.~B{\'e}rut, B.~Metzger, and Y.~Forterre, Revealing the frictional
  transition in shear-thickening suspensions, \emph{Proceedings of the National
  Academy of Sciences}. {\bf 114}\penalty0 (20), \penalty0 5147--5152,  (2017).

\bibitem{comtet2017pairwise}
J.~Comtet, G.~Chatt{\'e}, A.~Nigu{\`e}s, L.~Bocquet, A.~Siria, and A.~Colin,
  Pairwise frictional profile between particles determines discontinuous shear
  thickening transition in non-colloidal suspensions, \emph{Nature
  communications}. {\bf 8}\penalty0 (1), \penalty0 1--7,  (2017).

\bibitem{RamaswamyGLSKGSCCnn}
M.~Ramaswamy, I.~Griniasty, D.~B. Liarte, A.~Shetty, E.~Katifori, E.~D. Gado,
  J.~P. Sethna, B.~Chakraborty, and I.~Cohen.
\newblock Universal scaling of shear thickening transitions.
\newblock https://arxiv.org/abs/2107.13338,  (2021).

\bibitem{RamaswamyGSCCnn}
M.~{Ramaswamy}, I.~{Griniasty}, J.~P. {Sethna}, B.~{Chakraborty}, and
  I.~{Cohen}, {Incorporating tunability into a universal scaling framework for
  shear thickening}, \emph{arXiv e-prints}. art. arXiv:2205.02184 (May, 2022).

\bibitem{Trulsson16}
M.~Trulsson, E.~DeGiuli, and M.~Wyart, Effect of friction on dense suspension
  flows of hard particles, \emph{arXiv preprint arXiv:1606.07650}.  (2016).

\bibitem{ramaswamy2021universal}
M.~Ramaswamy, I.~Griniasty, D.~B. Liarte, A.~Shetty, E.~Katifori, E.~Del~Gado,
  J.~P. Sethna, B.~Chakraborty, and I.~Cohen, Universal scaling of shear
  thickening transitions, \emph{arXiv preprint arXiv:2107.13338}.  (2021).

\bibitem{DeGiuli17}
E.~DeGiuli and M.~Wyart, Unifying suspension and granular flows near jamming,
  \emph{EPJ Web Conf.} {\bf 140}, \penalty0 01003,  (2017).
\newblock URL \url{https://doi.org/10.1051/epjconf/201714001003}.

\bibitem{zheng1998conditions}
G.~Zheng and J.~R. Rice, Conditions under which velocity-weakening friction
  allows a self-healing versus a cracklike mode of rupture, \emph{Bulletin of
  the Seismological Society of America}. {\bf 88}\penalty0 (6), \penalty0
  1466--1483,  (1998).

\bibitem{Schwarz2001}
J.~Schwarz and D.~Fisher, Depinning with {{Dynamic Stress Overshoots}}: {{Mean
  Field Theory}}, \emph{Phys. Rev. Lett.} {\bf 87}\penalty0 (9), \penalty0
  096107,  (2001).
\newblock \doi{10.1103/PhysRevLett.87.096107}.

\bibitem{nicolas2016effects}
A.~Nicolas, J.-L. Barrat, and J.~Rottler, Effects of inertia on the
  steady-shear rheology of disordered solids, \emph{Physical review letters}.
  {\bf 116}\penalty0 (5), \penalty0 058303,  (2016).

\bibitem{de2022scaling}
T.~W. de~Geus and M.~Wyart, Scaling theory for the statistics of slip at
  frictional interfaces, \emph{arXiv preprint arXiv:2204.02795}.  (2022).

\bibitem{ozawa2021rare}
M.~Ozawa, L.~Berthier, G.~Biroli, and G.~Tarjus, Rare events and disorder
  control the brittle yielding of amorphous solids, \emph{arXiv preprint
  arXiv:2102.05846}.  (2021).

\bibitem{Popovic2018}
M.~Popovi{\' c}, T.~W.~J. de~Geus, and M.~Wyart, Elastoplastic description of
  sudden failure in athermal amorphous materials during quasistatic loading,
  \emph{Physical Review E}. {\bf 98}\penalty0 (4) (Oct, 2018).
\newblock ISSN 2470-0045, 2470-0053.
\newblock \doi{10.1103/PhysRevE.98.040901}.
\newblock URL \url{https://link.aps.org/doi/10.1103/PhysRevE.98.040901}.

\bibitem{Fielding2021}
S.~M. Fielding.
\newblock Yielding, shear banding and brittle failure of amorphous materials,
  (2021).

\bibitem{richard2021finite}
D.~Richard, C.~Rainone, and E.~Lerner, Finite-size study of the athermal
  quasistatic yielding transition in structural glasses, \emph{The Journal of
  Chemical Physics}. {\bf 155}\penalty0 (5), \penalty0 056101,  (2021).

\bibitem{chacko2021elastoplasticity}
R.~N. Chacko, F.~P. Landes, G.~Biroli, O.~Dauchot, A.~J. Liu, and D.~R.
  Reichman, Elastoplasticity mediates dynamical heterogeneity below the mode
  coupling temperature, \emph{Physical Review Letters}. {\bf 127}\penalty0 (4),
  \penalty0 048002,  (2021).

\end{thebibliography}


\end{document}